\documentclass[twocolumn]{aastex701}
\usepackage{amssymb,amsmath,graphicx,longtable,subfigure,wrapfig}
\usepackage{rotating}
\usepackage{float}
\usepackage{color}
\usepackage{dcolumn}

\shorttitle{The superMIGHTEE survey}
\shortauthors{Lal et~al.}

\begin{document}
\linenumbers

\title{The superMIGHTEE Project: MeerKAT and GMRT Together to Unveil the Deep Radio Sky}

\correspondingauthor{Dharam V. Lal}
\email{dharam@ncra.tifr.res.in}

\author[0000-0001-5470-305X]{Dharam V. Lal}
\email{dharam@ncra.tifr.res.in}
\affiliation{National Centre for Radio Astrophysics - Tata Institute of Fundamental Research Post Box 3, Ganeshkhind P.O., Pune 411007, India}

\author[0000-0001-0123-7890]{Russ Taylor}
\email{russ@idia.ac.za}
\affiliation{Inter-University Institute for Data Intensive Astronomy, Department of Astronomy, University of Cape Town, Private Bag X3, Rondebosch, 7701, Cape Town, South Africa}
\affiliation{Inter-University Institute for Data Intensive Astronomy, Department of Physics and Astronomy, University of the Western Cape, Robert Sobukwe Road, Bellville, 7535, Cape Town, South Africa}

\author[0000-0002-8418-9001]{Srikrishna Sekhar}
\email{srikrishna@idia.ac.za}
\affiliation{Inter-University Institute for Data Intensive Astronomy, Department of Astronomy, University of Cape Town, Private Bag X3, Rondebosch, 7701, Cape Town, South Africa}
\affiliation{National Radio Astronomy Observatory, NRAO, 1003 Lopezville Road, Socorro, NM 87801, USA}

\author[0000-0001-5356-1221]{Ch. Ishwara-Chandra}
\email{ishwar@ncra.tifr.res.in}
\affiliation{National Centre for Radio Astrophysics - Tata Institute of Fundamental Research Post Box 3, Ganeshkhind P.O., Pune 411007, India}

\author[0000-0002-6542-2939]{Sushant Dutta}
\email{sushant@idia.ac.za}
\affiliation{Inter-University Institute for Data Intensive Astronomy, Department of Astronomy, University of Cape Town, Private Bag X3, Rondebosch, 7701, Cape Town, South Africa}

\author[0000-0001-9821-4987]{Sthabile Kolwa}
\email{sthabile@idia.ac.za}
\affiliation{Inter-University Institute for Data Intensive Astronomy, Department of Astronomy, University of Cape Town, Private Bag X3, Rondebosch, 7701, Cape Town, South Africa}
\affiliation{Department of Mathematical Sciences, University of South Africa, Cnr Christian de Wet Rd and Pioneer Avenue, Florida Park, 1709, Roodepoort, South Africa}

\begin{abstract}
An international team of researchers has come together to undertake an ultra-broadband exploration of the deep radio sky.
The superMIGHTEE project combines data from the MIGHTEE project, using the precursor Square Kilometre Array (SKA) MeerKAT telescope in South Africa, with observations from the upgraded Giant Metrewave Radio Telescope (uGMRT) in India to produce deep images at several $\mu$Jy sensitivity over a frequency range of 200 MHz--2.5 GHz, with an angular resolution of a few arcseconds.
This paper describes the initial superMIGHTEE uGMRT data release, comprising total intensity continuum images covering a total of 9.9 deg$^2$ at 650 MHz and 6.9 deg$^2$ at 400 MHz in the XMM-LSS, COSMOS, and E-CDFS deep fields. The associated radio source catalogs include 27,101 sources at 650 MHz and 10,946 sources at 400 MHz. The redshift distribution of the sources extends to $z\sim4$ with a median value of $z=1$.
An overview of the broadband spectra of the sources, in combination with the MeerKAT MIGHTEE 1280 MHz data, reveals a clear change in spectral properties at the transition from an active galactic nuclei-dominated population to a population dominated by star-forming galaxies at flux densities of a few mJy. At higher frequencies, the star-forming galaxy population exhibits an optically thin synchrotron spectral index indicative of energy injection from supernovae. At lower frequencies, the spectra flatten significantly with decreasing flux density, and the fraction of sources with peaked spectra increases.
This is the first superMIGHTEE uGMRT data release. Subsequent releases will include spectropolarimetric and spectral line image cubes, as well as images at lower frequencies. The goal of the superMIGHTEE ultra-wideband dataset is to enhance our understanding of the evolution of active galactic nuclei and star-forming galaxies over cosmic time, shed light on the evolution of neutral hydrogen, and explore the origins and evolution of cosmic magnetic fields in clusters, filaments, and galaxies.

\end{abstract}

\keywords{Active galactic nuclei (16); Astrophysical black holes (98); Galaxy evolution (594); High-redshift galaxies (734); Intracluster medium (858); Radio continuum emission (1340); Radio galaxies (1343); Redshift surveys (1378); Star formation (1569);  Supermassive black holes (1663); X-ray active galactic nuclei (2035); Radio active galactic nuclei (2134)}
\section{Introduction}
\label{sec1}

The first decades of this century have seen tremendous advances in digital and information technologies, significantly impacting scientific inquiry. These advancements have been harnessed by the global radio astronomy community in the design of the Square Kilometre Array (SKA), a next-generation radio telescope array that began construction in Australia and South Africa this decade. The new technologies enabling the SKA have also led to significant improvements in the capabilities of existing radio telescopes and have fostered international collaboration on SKA pathfinder science programs. These programs help shape the scientific direction and address the technical challenges of the key science objectives of the SKA.

A deep survey covering tens of degrees of sky is expected to detect and catalog $\approx$10$^6$ galaxies, including typical star-forming galaxies in the nearby ($z$ $<$ 1) Universe, powerful starbursts at even greater redshifts, and active galactic nuclei (AGN) at the edge of the Universe; 
for example, deep ASKAP EMU survey of the GAMA23 field \citep{2022MNRAS.512.6104G}, 
the LOFAR two-meter Sky Survey \citep{2021A&A...648A...2S},
a wide-area GMRT 610-MHz survey of ELAIS~N1 field \citep{2020MNRAS.497.5383I}
the VLA-COSMOS 3 GHz large project \citep{2017A&A...602A...3D}, 
the sub-mJy radio sky in the extended Chandra Deep Field-South \citep{2013MNRAS.436.3759B},
deep multifrequency radio imaging in the Lockman Hole using the GMRT and VLA \citep{2009MNRAS.397..281I}, etc.
A global team of researchers, representing most of the countries participating in the SKA project, is working on a joint program to exploit these capabilities in exploring the deep radio sky as a scientific and technical pathfinder for the combined mid- and low-frequency SKA.
Below, we introduce our superMIGHTEE project, which integrates the precursor SKA MeerKAT telescope in South Africa and the SKA Pathfinder, the upgraded Giant Metrewave Radio Telescope (uGMRT) in India, to exploit their capabilities and create deep radio images.

The superMIGHTEE project had its genesis as a collaboration established under the bilateral India-South Africa Flagship Program in Astronomy, supported by the Department of Science and Technology in India and the Department of Science and Innovation in South Africa.
In South Africa, the MeerKAT International GHz Tiered Extragalactic Exploration (MIGHTEE) project \citep{Jarvis16,Hale_2024} is one of a small number of SKA precursor large survey programs being conducted with the MeerKAT telescope.
The superMIGHTEE project augments the MIGHTEE initiative by incorporating observations with the uGMRT at 125--250 MHz (band 2), 250--500 MHz (band 3), and 550--850 MHz (band 4), thereby providing over an octave of semicontinuous frequency coverage from 200 MHz to 2.5 GHz, with an angular resolution of a few arcseconds and \textsc{rms} sensitivities at the $\mu$Jy level.

\begin{table}
\begin{center}
\caption{A Summary of the uGMRT Observations of the MIGHTEE Fields to Date}
\begin{tabular}{l|ccccc|}
  \hline
    Survey  & uGMRT  & No. of Pointings &  Obs. Time\\ 
    Field   & Band &       & (hr) \\
  \hline
  XMM-LSS   & Band 3 &  4 & 36  &  \\ 
  XMM-LSS   & Band 4 & 19 &  145  \\ 
  E-CDFS    & Band 4 &  3 &   24 \\
  COSMOS    & Band 4 & 8  & 48 \\
  \hline
\end{tabular}
\label{tab:tab-time}
\end{center}
\end{table}

The paper is organized as follows:
We begin by highlighting the capabilities of the MeerKAT and uGMRT facilities in Section~\ref{sec:uGMRT-MeerKAT}.
Our observations, and data reduction, and analysis are summarized in Section~\ref{sec:obs}.
Section~\ref{sec:data-products}, and ~\ref{sec:src-astro-flux} describe the superMIGHTEE data products along with a few illustrations and validation of the contents of the superMIGHTEE Data Release~1 (DR1); e.g., source finding (Section~\ref{sec:data-products}), astrometry of our catalogs (Section~\ref{sec:5.1}), and flux density scale (Section~\ref{sec:5.2}).
The consensus photometric redshifts of our data products are presented in Section~\ref{sec:redshift}, and the early science results from our superMIGHTEE project, the spectral properties of the $\mu$Jy source population, are presented in Section~\ref{sec:spec-properties}.
Sections~\ref{sec:discussion} and \ref{sec:summary} present the scientific impact of our deep, ultra-broadband, high-sensitivity, high-resolution superMIGHTEE project, as well as a summary of our project, results, and future prospects, respectively.

Throughout this paper, we adopt a $\Lambda$CDM cosmology with $H_0$ = 70 km s$^{-1}$ Mpc$^{-1}$, $\Omega_{\rm m}$ = 0.27, and $\Omega_{\Lambda}$ = 0.73. We define the spectral index $\alpha$ as $S_\nu \propto \nu^\alpha$, where $S_\nu$ is the flux density at frequency $\nu$. The positions are given in J2000 coordinates.

\section{MeerKAT and the upgraded GMRT: Complementary Facilities}
\label{sec:uGMRT-MeerKAT}

The GMRT \citep{1991CSci...60...95S}, located on the Deccan Plateau in India, is the largest radio dish array in the world.
It has a hybrid configuration; 14 of its 30 antennas, each with a diameter of 45 m, are located in a central compact array approximately 1~km in size ($\simeq$ 5 k$\lambda$ at 1420 MHz), while the remaining antennas are distributed along three arms in a Y-shaped configuration, providing a maximum baseline of $\sim$ 25 km ($\simeq$ 120 k$\lambda$ at 1420 MHz).
Recently upgraded as an SKA Pathfinder, the GMRT now features a completely new set of broadband, highly sensitive receivers, a broadband digital correlator backend, and enhancements in its servo, mechanical, and electrical systems.
This upgraded GMRT, known as the uGMRT \citep{2017CSci..113..707G}, provides (nearly) seamless frequency coverage from 125 MHz to 1.5 GHz.

\begin{figure*}[ht]
\begin{center}
\includegraphics[width=0.98\textwidth]{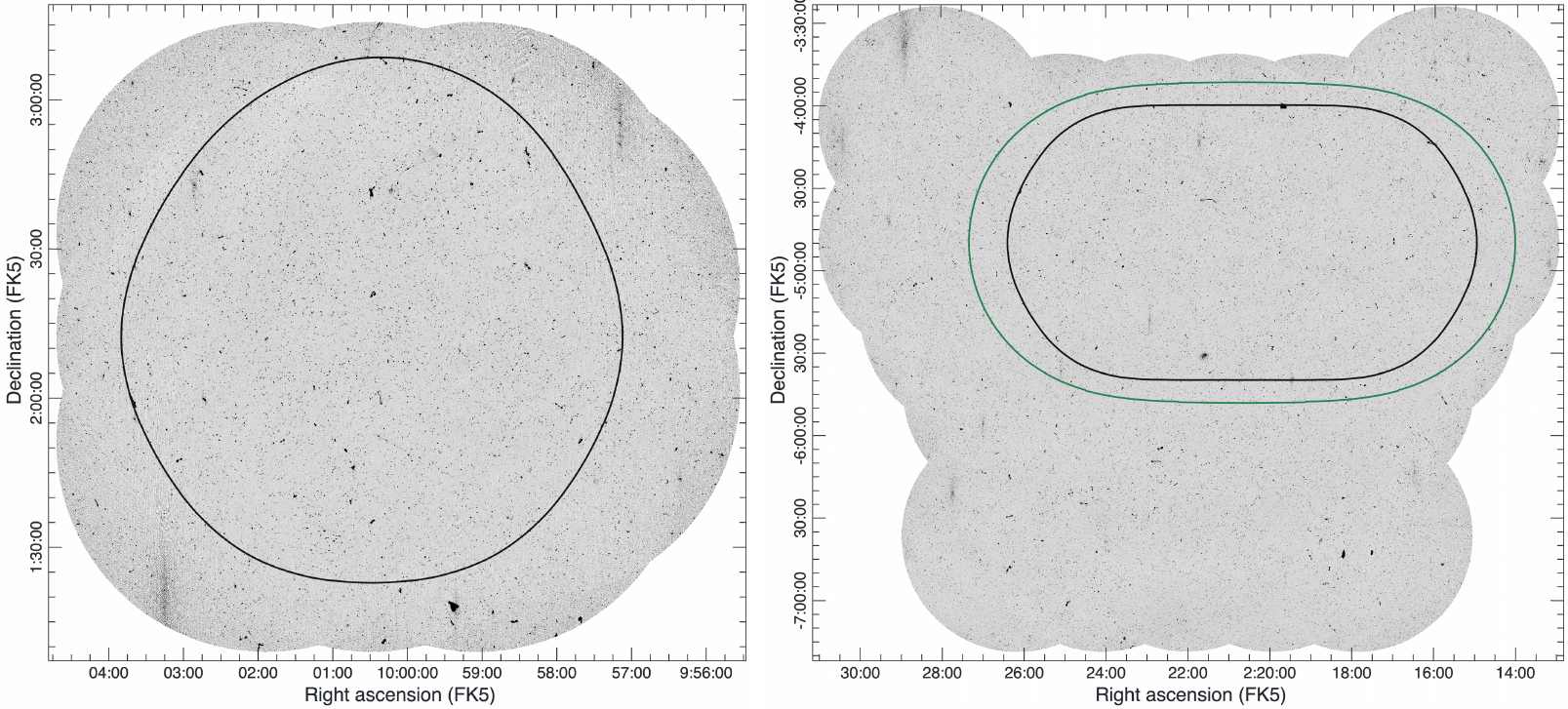}
\end{center}
\caption{The coverage of the COSMOS (left) and XMM-LSS (right) uGMRT observations superposed on the  MIGHTEE L-band (856--1711 MHz) mosaics. The green is band-3 (320--480 MHz) coverage, and the black is band-4 (561--740 MHz).}
\label{fig:coverage}
\end{figure*}

In South Africa, the MeerKAT radio telescope, originally called the Karoo Array Telescope \citep{2016mks..confE...1J}, serves as a precursor to the SKA mid-frequency dish array (SKA1–mid).
Completed in 2018, MeerKAT consists of 64 dishes, each 13.5 m in diameter.
Currently operated by the South African Radio Astronomy Observatory, it will later be integrated into the first phase of the SKA mid-frequency array.
MeerKAT is conducting several large-scale science programs addressing key SKA science objectives.
The telescope has three observing bands: UHF (544--1087 MHz), L-band (856--1711 MHz), and S-band (1750--3499 MHz).
Its array configuration is centrally condensed, with approximately 39 dishes within a 1 km radius ($\simeq$ 5 k$\lambda$ at 1420 MHz), while the remaining dishes are spread across an $\sim$ 8 km radius ($\simeq$ 40 k$\lambda$ at 1420 MHz).
This configuration provides MeerKAT with excellent surface brightness sensitivity and enables the generation of radio images with exceptionally high dynamic range and image fidelity.

Thus, MeerKAT and the uGMRT together offer tremendous synergy as complementary facilities for studying the deep radio sky.
The GMRT array has baselines nearly four times longer than those of MeerKAT, giving it a comparable imaging angular resolution at 250--850 MHz to that of MeerKAT at 1--3 GHz.
As a result, a single 8~hr synthesis observation with these instruments provides comprehensive coverage of the ($u,v$) plane, sampling both short and long baselines effectively.
This enables high angular resolution imaging ($\sim$5$^{\prime\prime}$) while maintaining very good sensitivity (a few $\mu$Jy~beam$^{-1}$) for detailed mapping of source structures.
Together, these two facilities can produce the most sensitive and scientifically powerful images of deep-sky radio emission before the SKA becomes operational.
The very broad combined bandwidth, spanning from $\sim$200 MHz to 2.5 GHz, provides a unique dataset that enables scientific investigations beyond the capabilities of any single instrument.
In the future, a similar spectral coverage will be achievable only by combining observations from the SKA-low facility in Australia and SKA-mid in South Africa.

\section{Observations and Data Processing}
\label{sec:obs}

\subsection{Observations}

Table~\ref{tab:tab-time} summarizes the observations of the uGMRT and MIGHTEE fields performed to date.
The MeerKAT MIGHTEE project has a total allocated observing time of 1920 hr, distributed across four extragalactic deep fields (E-CDFS, COSMOS, XMM-LSS, and ELAIS-S1), covering approximately 20 deg$^2$.
MIGHTEE observes at the L-band over an instantaneous 800 MHz bandwidth (880--1680 MHz), reaching an \textsc{rms} of $\sim$2\,$\mu$Jy, and achieving an angular resolution of $\sim$6$^{\prime\prime}$ (see also Table~\ref{tab:tab-time}).
MIGHTEE also obtains an ultra-deep image of a single pointing in E-CDFS, covering about 1 deg$^2$, reaching an \textsc{rms} noise level of $\sim$0.1\,$\mu$Jy through commensal observations with the MeerKAT LADUMA (Looking At the Distant Universe with the MeerKAT Array) large project \citep{2018AAS...23123107B}.
Additionally, we have secure S-band (1750--2750 MHz) data for a subregion of the MIGHTEE fields.
The L-band observations of all MIGHTEE fields using MeerKAT are complete, while the S-band observations are currently at various stages, including ongoing observations and data reduction.

\begin{table*}
\centering
\caption{The uGMRT Observations Analysed in This Data Release and Properties of Stokes-$I$ Continuum Mosaic Images}
\begin{tabular}{c|lcccclcc}
  \hline
    Field & \multicolumn{1}{c}{Obs\_ID} & Solid &  Obs. & Central & \multicolumn{1}{c}{Robust} &\hfil Synthesized \hfil & Noise & No. of \\
    & & Angle &  Band & Frequency && \hfil Beam \hfil &min/prob & Sources \\
    & & (deg$^2$) & &  (MHz) && \multicolumn{1}{c}{(maj $\times$ \,min, \, \,P.A.)} & ($\mu$Jy~beam$^{-1}$) \\
    \multicolumn{1}{c|}{(1)} & \multicolumn{1}{c}{(2)} & (3) & (4) & (5) & (6) & \multicolumn{1}{c}{(7)} & (8) & (9) \\
  \hline
XMM-LSS  & 36\_010 & 6.87   &  Band-3 & 400 & $-$0.4 & 6.9$^{\prime\prime}$ $\times$ \, 4.6$^{\prime\prime}$, \quad 64$^{\circ}$ &  17.2 / 32.0 & 10,931 \\ 
         &  & 6.87   &  Band-3 & 400 & $-$0.4 & 6.9$^{\prime\prime}$ $\times$ \, 6.9$^{\prime\prime}$              \hfil &  18.9 / 39.0 &  9,771 \\ 
         &  & 6.87   &  Band-3 & 400 & ~~0.0 & 7.7$^{\prime\prime}$ $\times$ \, 5.1$^{\prime\prime}$, \quad 61$^{\circ}$  &  16.0 / 32.5 &  10,946 \\ 
         &  & 6.87   &  Band-3 & 400 & ~~0.0 & 7.7$^{\prime\prime}$ $\times$ \, 7.7$^{\prime\prime}$              \hfil  &  17.1 / 38.3 &  9,745 \\ 
XMM-LSS  & 40\_044, 37\_031 & 5.02   &  band-4 & 650 & $-$0.4 & 5.0$^{\prime\prime}$ $\times$ \, 4.7$^{\prime\prime}$, \quad 52$^{\circ}$ &   4.6 / 8.3 & 16,284 \\ 
         &  & 5.02   &  Band-4 & 650 & $-$0.4 & 5.0$^{\prime\prime}$ $\times$ \, 5.0$^{\prime\prime}$             \hfill &   4.6 / 8.4 & 15,558 \\ 
         &  & 5.02   &  Band-4 & 650 & ~~0.0 & 5.7$^{\prime\prime}$ $\times$ \, 5.4$^{\prime\prime}$, \quad 49$^{\circ}$  &   4.6 / 9.4 & 13,859 \\ 
         &  & 5.02   &  Band-4 & 650 & ~~0.0 & 5.7$^{\prime\prime}$ $\times$ \, 5.7$^{\prime\prime}$             \hfill  &  4.6 / 9.6 & 13,259 \\ 
 COSMOS  & 41\_094 & 2.94   &  band-4 & 650 & $-$0.4 & 5.7$^{\prime\prime}$ $\times$ \, 5.1$^{\prime\prime}$, \quad 71$^{\circ}$ &  5.0 / 8.8 &  7,530 \\
         &  & 2.94   &  Band-4 & 650 & $-$0.4 & 5.7$^{\prime\prime}$ $\times$ \, 5.7$^{\prime\prime}$              \hfil &  5.0 / 10.1 &  7,100 \\
         &  & 2.94   &  Band-4 & 650 & ~~0.0 & 6.6$^{\prime\prime}$ $\times$ \, 6.0$^{\prime\prime}$, \quad 70$^{\circ}$  &  4.3 / 11.2 &  7,225 \\
         &  & 2.94   &  Band-4 & 650 & ~~0.0 & 6.6$^{\prime\prime}$ $\times$ \, 6.6$^{\prime\prime}$              \hfil  &  4.1 / 11.7 &  6,785 \\
 E-CDFS  & 33\_055 & 1.91   &  band-4 & 650  & $-$0.4  &  5.0$^{\prime\prime}$ $\times$ \, 3.5$^{\prime\prime}$, \quad 28$^{\circ}$   & 8.8 / 21.8 & 3,249 \\ 
         &  & 1.91   &  Band-4 & 650  & $-$0.4  &  5.0$^{\prime\prime}$ $\times$ \, 5.0$^{\prime\prime}$                 &  9.5 / 26.5 & 2,580 \\ 
         &  & 1.91   &  Band-4 & 650  & ~~0.0  &  5.8$^{\prime\prime}$ $\times$ \, 4.2$^{\prime\prime}$, \quad 32$^{\circ}$   & 8.2 / 21.9 & 2,934 \\ 
         &  & 1.91    &  Band-4 & 650  & ~~0.0  &  5.8$^{\prime\prime}$ $\times$ \, 5.8$^{\prime\prime}$    & 8.5 / 29.5 & 2,282 \\ 
  \hline
\end{tabular}
\label{tab:mosaics}
\tablecomments{Column~(1): observed field name.
Column~(2): GMRT observation program code.
Column~(3): solid angle within each mosaic.
Columns~(4) and (5): uGMRT observing band and central frequency of the observing band.
Column~(6): Briggs robustness parameter.
Column~(7): synthesized beam (major $\times$ minor axes, position angle).
Column~(8): minimum and most probable \textsc{rms} values.
Column~(9): total number of sources detected.}
\end{table*}

The uGMRT data includes both new observations and archival data that are not part of the superMIGHTEE time allocation. 
All observations to date were conducted over three GMRT observing semesters, from 2019 September to 2022 November, using the wideband correlator backend.
These observations consist of several overlapping pointings within three MIGHTEE fields, XMM-LSS, E-CDFS, and COSMOS.
For example, in band 4, there are 19 pointings in XMM-LSS, eight in COSMOS, and four in E-CDFS.
The separation of pointings in band 4 and band 3 is 17.6$^\prime$ and 30.5$^\prime$, respectively, which corresponds to one-half the FWHM of the uGMRT primary beams at 800 MHz and 450 MHz.
Figure~\ref{fig:coverage} illustrates the coverage of the uGMRT mosaics in the COSMOS and XMM-LSS fields, superimposed on the MIGHTEE L-band mosaic images (see also Section~\ref{sec:data-products}).

Each pointing is observed during a long track of approximately 8 hr. To minimize ionospheric effects on polarimetry, most observations were conducted at night.
During each observation, we observed 3C\,48 and 3C\,147 for several minutes to perform flux density and bandpass calibration. We observed a polarized source (J0323$+$055 for XMM-LSS, J0943$-$083 for COSMOS, and J0240$-$231 for E-CDFS) every 40 minutes during the 8~hr tracks for time-dependent gain and polarization calibration. 

\subsection{Calibration and Imaging}

The MIGHTEE full-polarization observation, calibration, and imaging processes are described in \cite{Taylor_2024}.
The uGMRT data were processed with the same \textsc{CASA}-based, \textsc{processMeerKAT} pipeline infrastructure \citep{Collier_2021} on the IDIA ilifu data-intensive research cloud\footnote{\url{https://ilifu.ac.za}} as the MIGHTEE polarization processing, but modified for circularly polarized feeds. 

Briefly, the \textsc{processMeerKAT} pipeline partitions the full band into several spectral windows (SPWs) of approximately 30 MHz each. Each SPW is processed independently and concurrently, and all SPWs are concatenated into a single measurement set (\textsc{ms}) at the end of a priori calibration, before self-calibration and further processing.
The band-3 observations were processed between 300\,MHz and 480\,MHz, while the band-4 observations were processed between 560\,MHz and 816\,MHz. The calibration procedures followed for both bands were broadly similar. However, the band-4 observations underwent a single round of flagging and calibration, whereas the band-3 observations required two rounds of each to account for the more active radio frequency interference (RFI) environment at lower frequencies.
The calibration procedure consists of the following steps:
\begin{figure}[t]
\begin{center}
\includegraphics[width=\columnwidth]{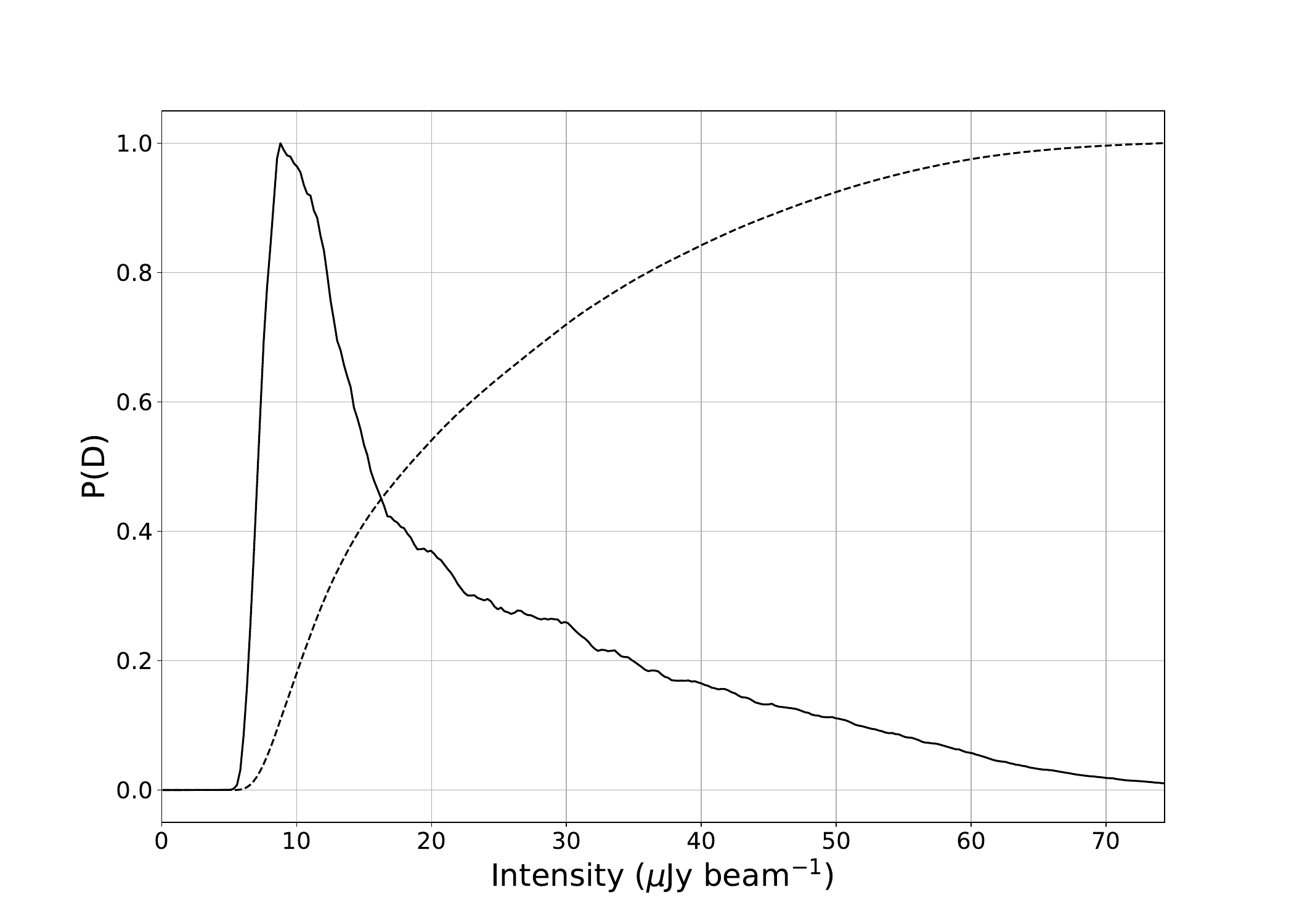}
\end{center}
\caption{The normalized probability density distribution (solid line) of the values in the \textsc{PyBDSF} \textsc{rms} image of the COSMOS band-4 mosaic.  The most probable value of the \textsc{rms} is 8.8 $\mu$Jy~beam$^{-1}$.  The dashed line shows the cumulative distribution.}
\label{fig:noise}
\end{figure}

\begin{itemize}
    \item[(i)] Following the initial partition stage, the data are flagged using the \textsc{CASA} task \textsc{flagdata}, using the \textsc{clip} and \textsc{tfcrop} flagging modes.
    \item[(ii)] Next, the a priori calibration is performed, i.e.,
    \begin{itemize}
        \item[$-$] First the calibration model is set using the \textsc{setjy} task \cite[using the][standard]{2017ApJS..230....7P}.
        \item[$-$] The bandpass shape is solved for using the primary calibrator, followed by time-dependent gain calibration on the primary, secondary, and polarization calibrators.
        \item[$-$] The cross-hand delay and R-L phase calibration are solved using on the polarization calibrator, while the leakages are solved using the (unpolarized) primary calibrator. 
        \item[$-$] After solving, all calibration solutions are applied, followed by a second round of flagging on the data.
        As noted above, calibration for band 4 data stops at this stage, and self-calibration begins. However, for band 3 data, all previous calibration solutions are cleared at this point, and the entire calibration process is repeated from the beginning. This additional step is necessary to prevent excessive RFI from contaminating the solutions, particularly because the signal in the cross-hand data can be weak, and the presence of RFI can significantly skew the polarization calibration.
    \end{itemize}
    \item[(iii)] Self-calibration was performed in a standard manner for both bands, i.e., two rounds of phase-only self-calibration were followed by two rounds of amplitude and phase self-calibration, with decreasing calibration solution interval from 10 minutes to 1 minute.
    \item[(iv)] This self-calibrated ($u,v$) data was imaged using the \textsc{tclean} task. During imaging, a ($u,v$) threshold of 0.5 k$\lambda$ was applied to eliminate the shortest baselines, which are most affected by RFI. This cut does not severely hamper diffuse source sensitivity, while improving the overall image \textsc{rms}. 
    \item[(v)] The \textsc{mt-mfs} images from each pointing were mosaiced together using the \textsc{casa} \textsc{linearmosaic} tool, which uses a linear mosaicing to create the primary beam weighted mosaic.  Wideband primary beam responses were constructed using the prescription for the frequency dependence of the primary beam, available at the GMRT website\footnote{\url{https://www.gmrt.ncra.tifr.res.in}} (dated 2023 November 29) for each field.
    In constructing the mosaics, each individual pointing was 
    weighted by the square of the inverse \textsc{rms} noise of that image. The \textsc{rms} noise was calculated using the  \textsc{CASA} task \textsc{imstat} with \textsc{algorithm=`fit-half'},
    which estimates the \textsc{rms} noise using the distribution of real and virtual pixel values, where the virtual part of the dataset is created by reflecting all the real values through the center value, in the image \citep[see also][]{2022PASP..134k4501C}.

    \item[(vi)] In addition to the broadband \textsc{mt-mfs} images, we also construct full-Stokes hypercubes using the IDIA cube generation  pipeline\footnote{\url{https://githup.com/idia-astro/frocc}}.
\end{itemize}

\section{Data Products}
\label{sec:data-products}

The primary image data products of the superMIGHTEE data products include broadband, high-sensitivity total intensity continuum images at 400 and 650\,MHz, spectropolarimetric hypercubes, and high-resolution total intensity cubes.
These data explore both the spectral energy distribution of total radio intensity and the spectral dependence of polarization.  This paper presents the first superMIGHTEE uGMRT DR1 that includes total intensity images and associated radio source catalogs. The spectropolarimetric hypercubes and high spectral resolution spectral line cubes will be the subject of subsequent data releases.

\begin{figure*}[t]
\begin{center}
\includegraphics[width=\textwidth]{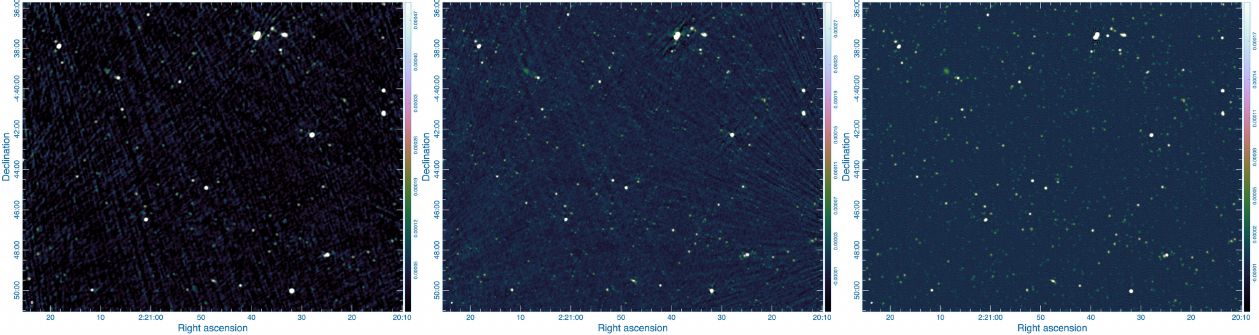}
\includegraphics[width=\textwidth]{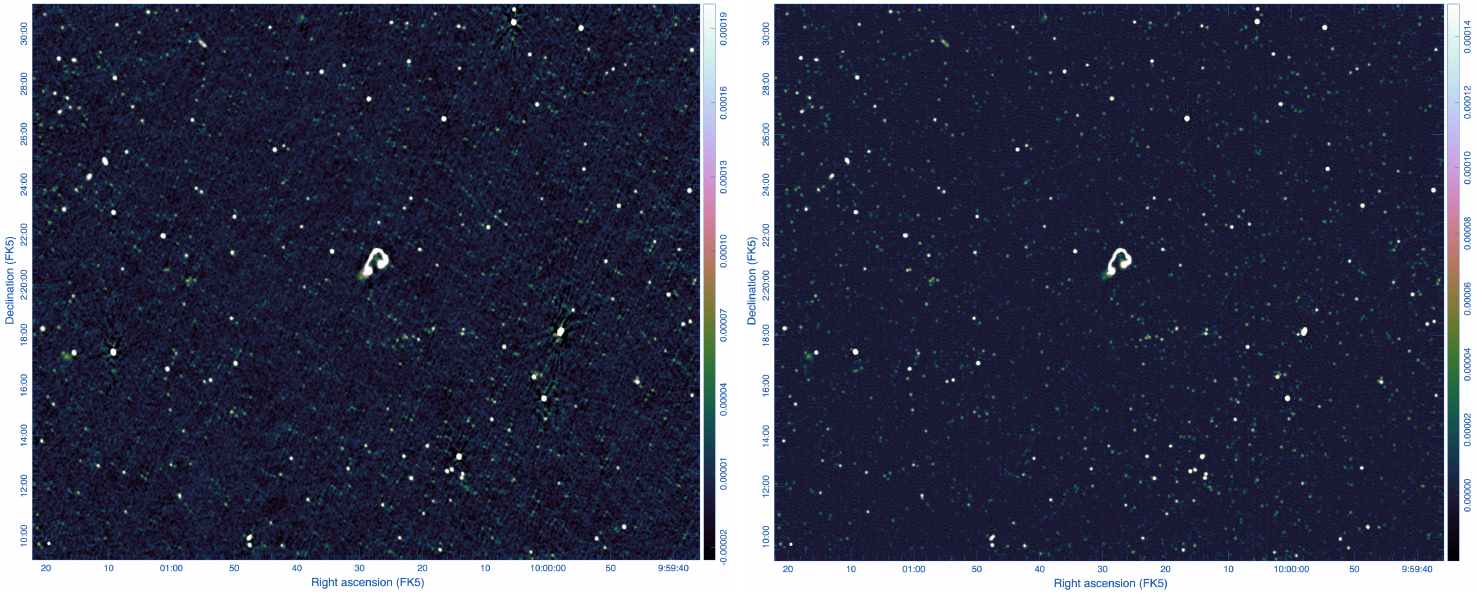}
\end{center}
\caption{Images of a region of the XMM-LSS (upper panel) field at band-3 (250--500 MHz)
at 6.9$^{\prime\prime}$ resolution (left), band-4 (550--850 MHz) at 5$^{\prime\prime}$ resolution (middle) and the MeerKAT MIGHTEE image at 1.284 GHz at 5$^{\prime\prime}$ resolution (right), and the COSMOS field (lower panel) at band-4 (550--850 MHz) at 5.6$^{\prime\prime}$ resolution (left) and the MeerKAT MIGHTEE image at 1.284 GHz at 5.2$^{\prime\prime}$ resolution (right).
The images were created using the \textsc{carta} software with a CubeHelix color mapping.}
\label{fig:b3-b4-xmmlss-cosmos}
\end{figure*}

Figure~\ref{fig:coverage} illustrates the coverage of uGMRT continuum images in the COSMOS and XMM-LSS fields. The lines represent the locations of the 0.04 mosaic weight for each of the band-4 (550--850 MHz) and band-3 (250--500 MHz) mosaic images.
The mosaic weights are given by the square of the wideband primary beam, so a weight of 0.04 corresponds to an equivalent noise level at the 20\% primary beam point in a single-pointing image. The continuum images are constructed using Briggs weighting with robust parameters of $-$0.4 and 0.0. Due to variations in $(u,v)$ coverage between pointings, the synthesized beam dimensions vary by $\lesssim$8\% between observations. Consequently, a raw mosaic of individual images results in a beam that varies across the mosaic. Thus, to construct a large mosaic with a uniform beam, two separate mosaics are created for each robust setting; one where individual pointings are smoothed to the minimum beam dimension that encompasses all the beams of the individual pointings, and another where the individual pointings are smoothed to the minimum circular beam.

Since the wideband primary beam changes with distance from the pointing center, the effective central frequency of the observation depends on position within the mosaic. Therefore, an effective frequency mosaic is generated for each mosaic by mosaicing effective frequency images for each pointing with the same mosaic weights as the target fields. The effective frequency mosaic image gives the effective frequency of the flux density at each pixel. Additionally, images of the mosaic weights are included as part of the superMIGHTEE data products.

\begin{figure*}[ht]
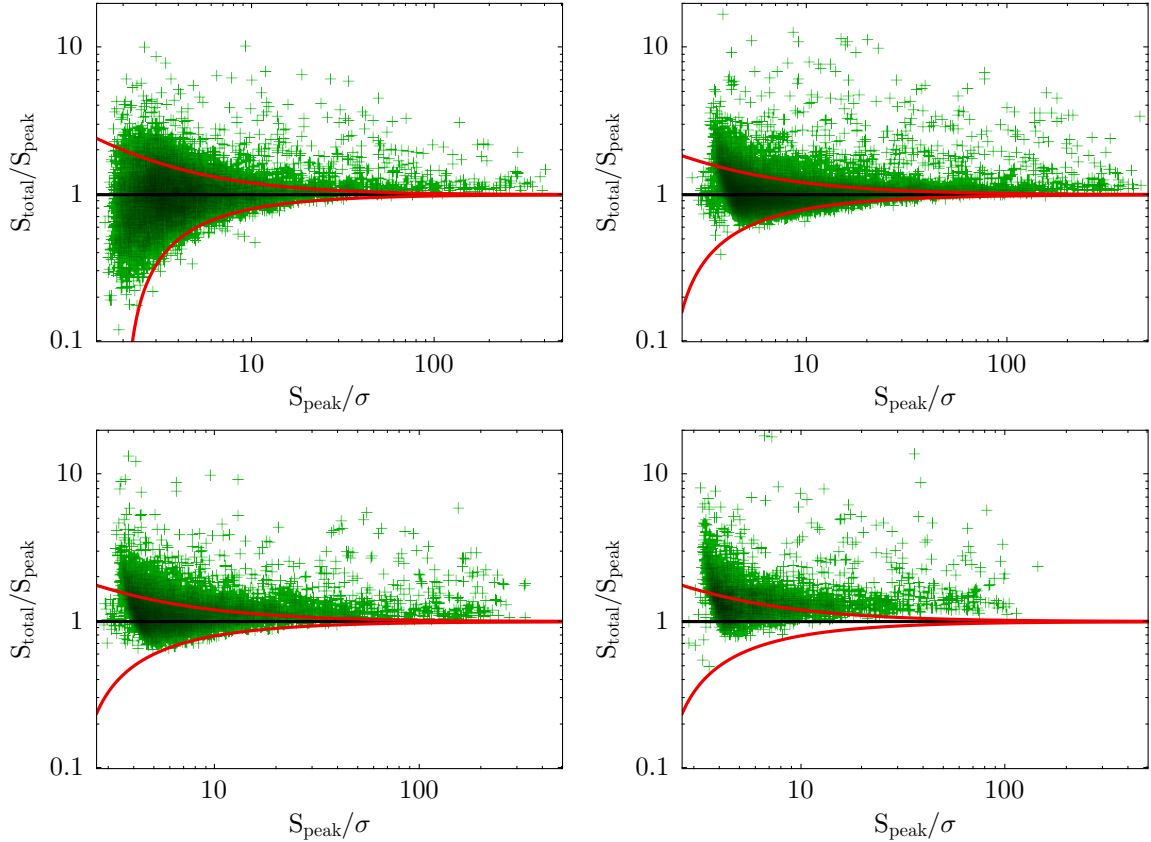

\begin{center}
\includegraphics[scale=0.4]{Flux_Ratio_XMMLSS_Band-3-r-0.4.png}
\includegraphics[scale=0.4]{Flux_Ratio_XMMLSS_Band-4-r-0.4.png}
\includegraphics[scale=0.4]{Flux_Ratio_COSMOS_Band-4-r-0.4.png}
\includegraphics[scale=0.4]{Flux_Ratio_ECDFS_Band-4-r-0.4.png}
\end{center}
\caption{Plots showing the ratio of total to peak flux density versus signal to noise.
The lower and upper envelopes, shown by the red solid lines with $A$ = 2.0 for all the fields (see also Section~\ref{sec:data-products}), encompass the unresolved sources in the samples.
The upper left, upper right, lower left, and lower right panels show results for XMM-LSS band 3, and XMM-LSS, COSMOS, and E-CDFS band 4, respectively, based on images for robust = $-0.4$.}
\label{fig:fluxratio}
\end{figure*}

Table~\ref{tab:mosaics} lists the properties of the uGMRT mosaic images for all bands and fields.
The columns are as follows: (1) field name; (2) GMRT observation program code; (3) solid angle within each mosaic; (4) uGMRT observing band; (5) central frequency of the observing band; (6) Briggs robustness parameter; (7) synthesized beam (major $\times$ minor axes, position angle); (8) minimum and most probable \textsc{rms} noise values; and (9) number of sources detected.
Note that the listed area corresponds to the solid angle within a mosaic weight of 0.04 for each mosaic image.
The image \textsc{RMS} is higher close to strong sources due to direction-dependent errors.
The close spacing of the mosaic pointings somewhat mitigates direction-dependent effects; however, direction-dependent calibration will be explored in detail in our second superMIGHTEE data release, which will also include observations of the COSMOS field in band 3.

A component source catalog was generated for each mosaic using the Python Blob Detection and Source Finder \textsc{PyBDSF\footnote{\url{https://github.com/lofar-astron/PyBDSF}}} source-finding software \citep{pybdsf}.
We applied ``forced photometry'' for the XMM-LSS band-3 images using source-fitting islands determined from the positions of the band-4 component catalog (robust = $-$0.4). 
In a mosaic image, noise varies according to the mosaic weights and is a factor of $\lesssim$ 3 higher at the outer edges, where weights are lower. 
During the preprocessing step, {\sc PyBDSF} creates a background root mean square (\textsc{rms}) map and a mean intensity image map.  An {\sc rmsbox} = (40,13) is used for the \textsc{rms} calculation.
Thus, the \textsc{rms} values listed in  Table~\ref{tab:mosaics} are the minimum and most probable values of the \textsc{rms} distribution from the \textsc{PyBDSF-rms} images (see Figure~\ref{fig:noise}). 
The upper panel of Figure~\ref{fig:b3-b4-xmmlss-cosmos} displays cutout sections of the XMM-LSS mosaic at band-3 and band-4 in a region free from strong sources, alongside a corresponding cutout from the MIGHTEE L-band image. 
Similarly, the lower panel of Figure~\ref{fig:b3-b4-xmmlss-cosmos} displays a cutout of the COSMOS band-4 image alongside a corresponding cutout from the MIGHTEE L-band image. 
Note that the uGMRT band-4 images exhibit similar detection levels for synchrotron sources and nearly identical resolution to the MIGHTEE L-band images.
The MIGHTEE DR1 5$^{\prime\prime}$ resolution images have central \textsc{rms} values of 2.4~$\mu$Jy~beam$^{-1}$ and 
3.6~$\mu$Jy~beam$^{-1}$ for COSMOS and XMM-LSS, respectively, at $\sim$1250\,MHz \citep{Hale25}. 
Assuming a spectral index of $-$0.75, the same signal-to-noise ratio is achieved with an \textsc{rms} of 4.0~$\mu$Jy~beam$^{-1}$ and 6.0~$\mu$Jy~beam$^{-1}$ for COSMOS and XMM-LSS, respectively, at 650~MHz, which is close to our values in Table~\ref{tab:mosaics}.

\begin{table}
\centering
\caption{Astrometric Corrections for Each Mosaic}
\label{tab-astrometry}
    \begin{tabular}{l c |  r r}
    \hline
    Field  & &  $\Delta \alpha$\,(arcsec)~~~~~& $\Delta \delta$\, (arcsec)~~~~~\\
    \hline
    XMM-LSS & (B3) & ~~~$0.051 \pm 0.009$ & ~~~$0.092 \pm 0.008$ \\
    XMM-LSS & (B4) &$0.028 \pm 0.003$ & $0.078 \pm 0.003$ \\
    COSMOS & (B4) &~~~$0.039 \pm 0.006$ & ~~~$-0.120 \pm 0.006$ \\
    E-CDFS & (B4) &~~~$-0.031 \pm 0.010$ & $0.009 \pm 0.010$ \\
    \hline
    \end{tabular}
\end{table}

Using the highest resolution common-beam mosaic images, in the Figure~\ref{fig:fluxratio}, we plot the ratio of total flux density ($S_{\rm total}$) to peak flux density ($S_{\rm peak}$) as a function of the signal-to-noise for all sources in the XMM-LSS, COSMOS, and E-CDFS robust $-$0.4 fields.

For the unresolved population, we expect scatter around $R = 1$ due to \textsc{rms} noise.
If the noise is normally distributed, 95.45\% of the unresolved sources will fall within $\pm2\sigma$ of unity.
The red curves on Figure~\ref{fig:fluxratio} are a functions of the form
$$
R = 1 \pm \frac{A}{x}
$$
where $x = S_{\rm peak}/\sigma$.
Setting $A$ = 2.0, the red curves encompass the $2\sigma$ scatter for unresolved sources (see Figure~\ref{fig:fluxratio}). 
The data points are systematically above unity for the stronger source population.
This is consistent with the known angular size -- flux density relationship \citep{Windhorst_1984}, which implies a median angular size of $\sim10^{\prime\prime}$ for sources $\gtrsim$500\,$\mu$Jy~beam$^{-1}$ at 1.4\,GHz.
For sources with lower signal-to-noise ratios, the $2\sigma$ curve provides a good fit to the lower envelope of the XMM-LSS and COSMOS distributions.  
It is clear that the majority of these fainter sources are unresolved.
In the XMM-LSS field, 85\% of the sources in band~3 and 82\% of the sources in band~4 fall within the $2\sigma$ envelope, while in the COSMOS field, 65\%  of the sources in band~4 lie within this range.
The E-CDFS field shows a more resolved population, with only 42\% of the sources in band~4 are falling within the $2\sigma$ boundaries.
This can be attributed to two factors: (i) The noise in E-CDFS is roughly twice that of XMM-LSS and COSMOS, leading to higher source flux densities, and thus the median angular sizes are larger. (ii) Additionally, E-CDFS has the highest angular resolution, with a beam solid angle 34\% smaller than that of XMM-LSS.

\section{Astrometry and Flux Density Scale}
\label{sec:src-astro-flux}

\subsection{Astrometric Precision}
\label{sec:5.1}

Systematic astrometric errors in superMIGHTEE radio mosaics are measured using the positions in \textsc{PyBDSF} source catalogs for the XMM-LSS, COSMOS, and E-CDFS fields. In order to correct these astrometric errors, we determine how offset the radio source positions are from the positions of their optical/near-infrared (NIR) counterparts, on average. For this, the superMIGHTEE radio source lists are crossmatched with optical and NIR catalogs based on surveys covering the target superMIGHTEE DR1 fields (see Section~\ref{sec:redshift}). The crossmatch radii used range between 1$^{\prime\prime}$ and 2$^{\prime\prime}$ depending on how the radio sources are distributed in projection, with more crowded fields requiring a lower crossmatch radius. The offsets in both Right Ascension (R.A./$\alpha$) and Declination (decl./$\delta$) between the positions of radio and optical/NIR counterparts are measured as the median offset in R.A.\ and decl.\ over the cross-matched samples. Table~\ref{tab-astrometry} lists the radio-optical astrometric offsets ($\Delta \alpha$\ and $\Delta \delta$ in arcseconds) for each of the superMIGHTEE mosaics. The image mosaics and the corresponding catalogs are corrected for these offsets. 

\begin{table*}
\centering
\caption{Radio Crossmatch Statistics between Frequency Bands}
\label{tab:supermightee-radio-statistics}
    \begin{tabular}{l l | c r r}
    \hline
    Field  & Crossmatched & Matched Fraction & Median $\Delta \alpha$ & Median $\Delta \delta$ \\
           & Observing Bands &     (\%)     & \multicolumn{1}{c}{(arcsec)} & \multicolumn{1}{c}{(arcsec)} \\
    \multicolumn{1}{c}{(1)} & \multicolumn{1}{c|}{(2)} & \multicolumn{1}{c}{(3)} & \multicolumn{1}{c}{(4)} & \multicolumn{1}{c}{(5)} \\
    \hline
    XMM-LSS & uGMRT B4-B3 & 10,152/16,843 (60.3 \%) & ~~~$0.040 \pm 0.001$ & ~~~$0.008 \pm 0.001$ \\
    XMM-LSS & uGMRT B4-MeerKAT & 14,014/16,843 (83.2 \%) & $-0.012 \pm 0.001$ & $-0.013 \pm 0.001$ \\
    COSMOS & uGMRT B4-MeerKAT & 6,617/7,340 (90.1 \%) & ~~~$0.083 \pm 0.001$ & ~~~$0140\pm 0.002$ \\
    E-CDFS & uGMRT B4-MeerKAT & 2,434/3,386 (71.9 \%) & ~~~$0.020 \pm 0.001$ & $-0.059 \pm 0.001$ \\
    \hline
    \end{tabular}
\tablecomments{Column~(1) is the observed field name. Column~(2) lists the survey field and the two radio frequency catalogs that have been crossmatched. Column~(3) shows the number of uGMRT band-4 sources recovered in the crossmatch and the percentage of sources that matched. Columns~(4) and (5) give the median offsets in R.A. and decl. between radio catalogs.}
\end{table*}

After astrometric correction, we checked the positional accuracy between the superMIGHTEE catalogs and the MIGHTEE catalogs. We further crossmatch uGMRT band-4 mosaic images with band-3 mosaic images and then with MIGHTEE DR1 images for the XMM-LSS field. Similarly, for the COSMOS and E-CDFS fields, we crossmatch uGMRT band-4 mosaic images with MIGHTEE DR1 images.

\begin{figure*}
    \centering
    \subfigure[XMM-LSS position offsets for uGMRT band-3 (B3) and band-4 (B4).]{\includegraphics[width=0.36\textwidth]{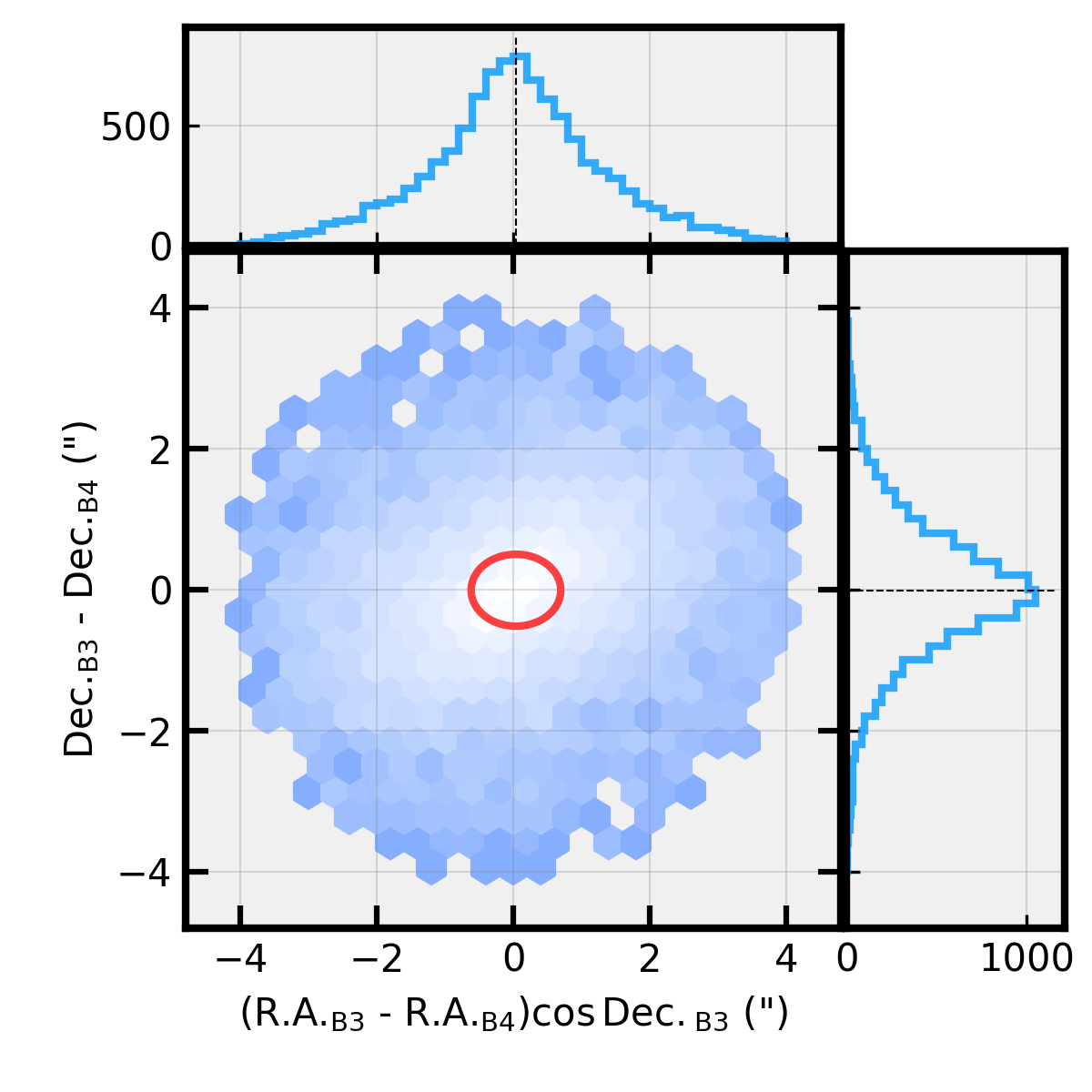}}
    \subfigure[XMM-LSS position offsets for uGMRT band-4 (B4) and MIGHTEE DR1 (MK).]{\includegraphics[width=0.36\textwidth]{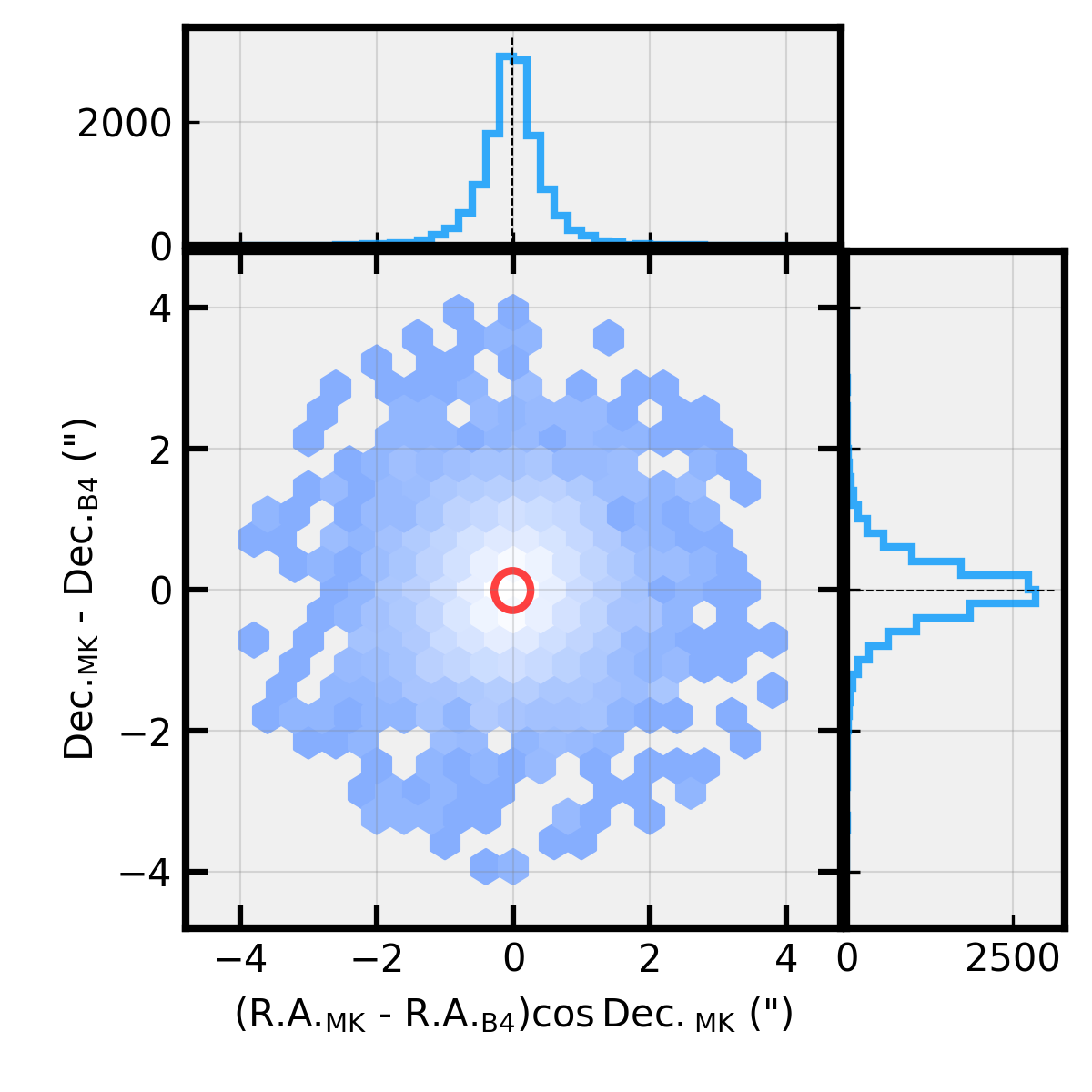}}
    \\
    \subfigure[COSMOS position offsets for uGMRT band-4 (B4) and MIGHTEE DR1 (MK).]{\includegraphics[width=0.35\textwidth]{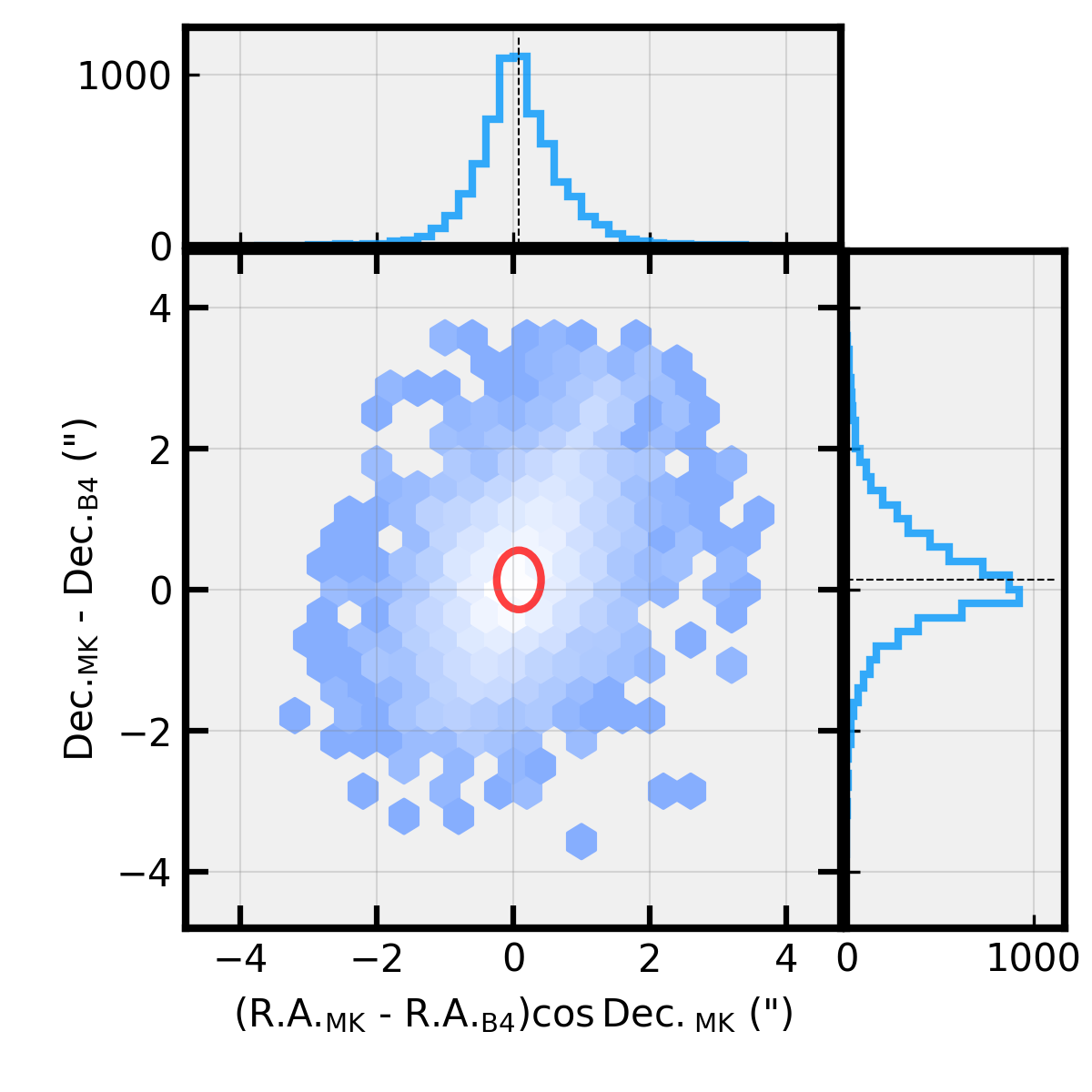}}
    \subfigure[E-CDFS position offsets between uGMRT band-4 and MIGHTEE DR1 (MK).]{\includegraphics[width=0.35\textwidth]{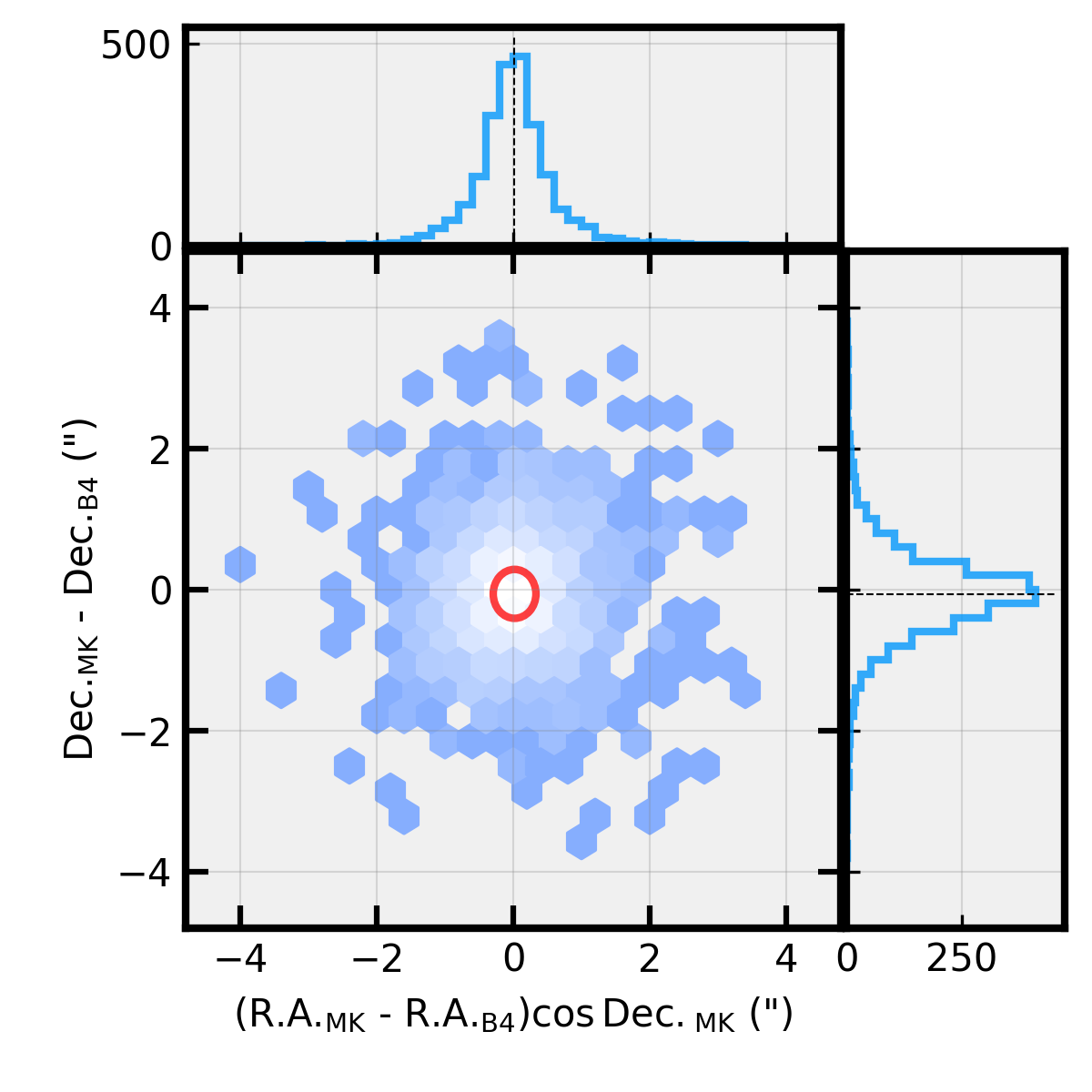}}
    \label{fig:supermightee-positional-accuracy}
    \caption{The positional accuracy of the radio catalogs after astrometry correction is represented for XMM-LSS, COSMOS, and E-CDFS fields. The enclosed region of the red ellipse in each plot represents the median offset and uncertainty in R.A. and decl.}
\end{figure*}

Table \ref{tab:supermightee-radio-statistics} lists the radio crossmatching statistics across the target fields, along with the median offsets in R.A. and decl. Figure \ref{fig:supermightee-positional-accuracy} shows the two-dimensional distributions of R.A. and decl. offsets for each of the observed fields. The XMM-LSS bands have astrometric consistency within 0.04$^{\prime\prime}$. The MIGHTEE DR1 data sets, however, are not corrected for astrometry against optical, and show differences with uGMRT at the level of up to several tenths of an arcsecond. These offsets are similar in magnitude to those observed for the MIGHTEE DR1 image sets by \cite{Hale_2024}.

\subsection{Flux Density Scale}
\label{sec:5.2}
To validate the flux density scale, we fit the radio spectral energy distribution (SED) of several prominent, bright radio sources within each mosaic field using a curved power-law model. 
The curved power-law model has the form 
$$
S_{\nu} = S_{0} \left(\frac{\nu}{\nu_{0}}\right)^{\alpha} \cdot e^{q(\ln \nu)^2};
$$
where the spectral index, $\alpha$, and the flux density normalization, $S_0$ are determined by fitting, and $\nu_{0}$ is the reference frequency at which $S_0$ is evaluated.  Here, $q$ parametrizes the spectral curvature. 
Specifically, $q<$ 0 results in a convex spectrum, and as 
$q$ approaches zero, the curvature diminishes and $\alpha$ becomes the
standard spectral index \citep[see][]{Callingham17,Quici21,Dutta23}.

\begin{figure*}[ht]
\begin{center}
\includegraphics[scale=0.05]{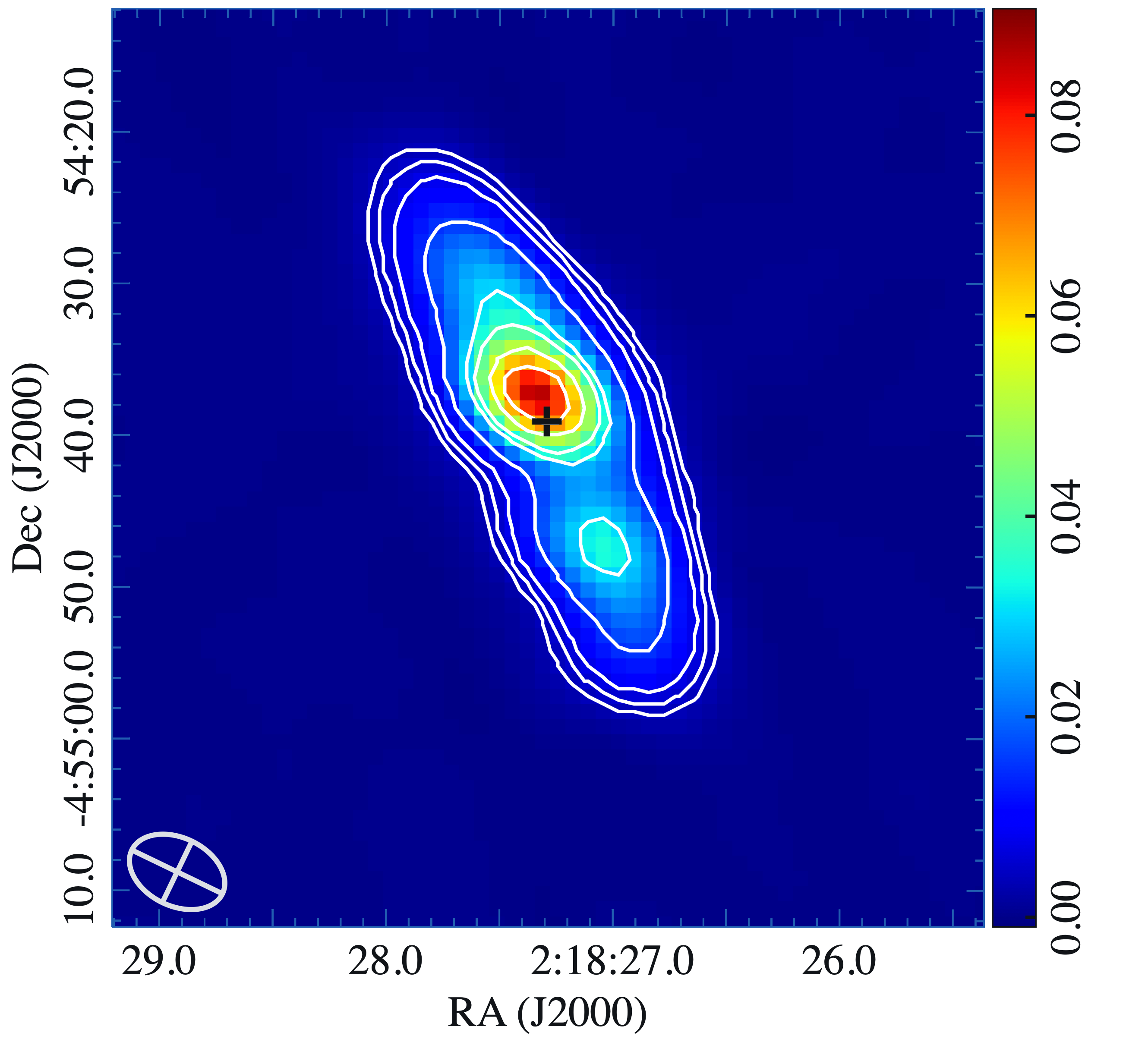}
\includegraphics[scale=0.05]{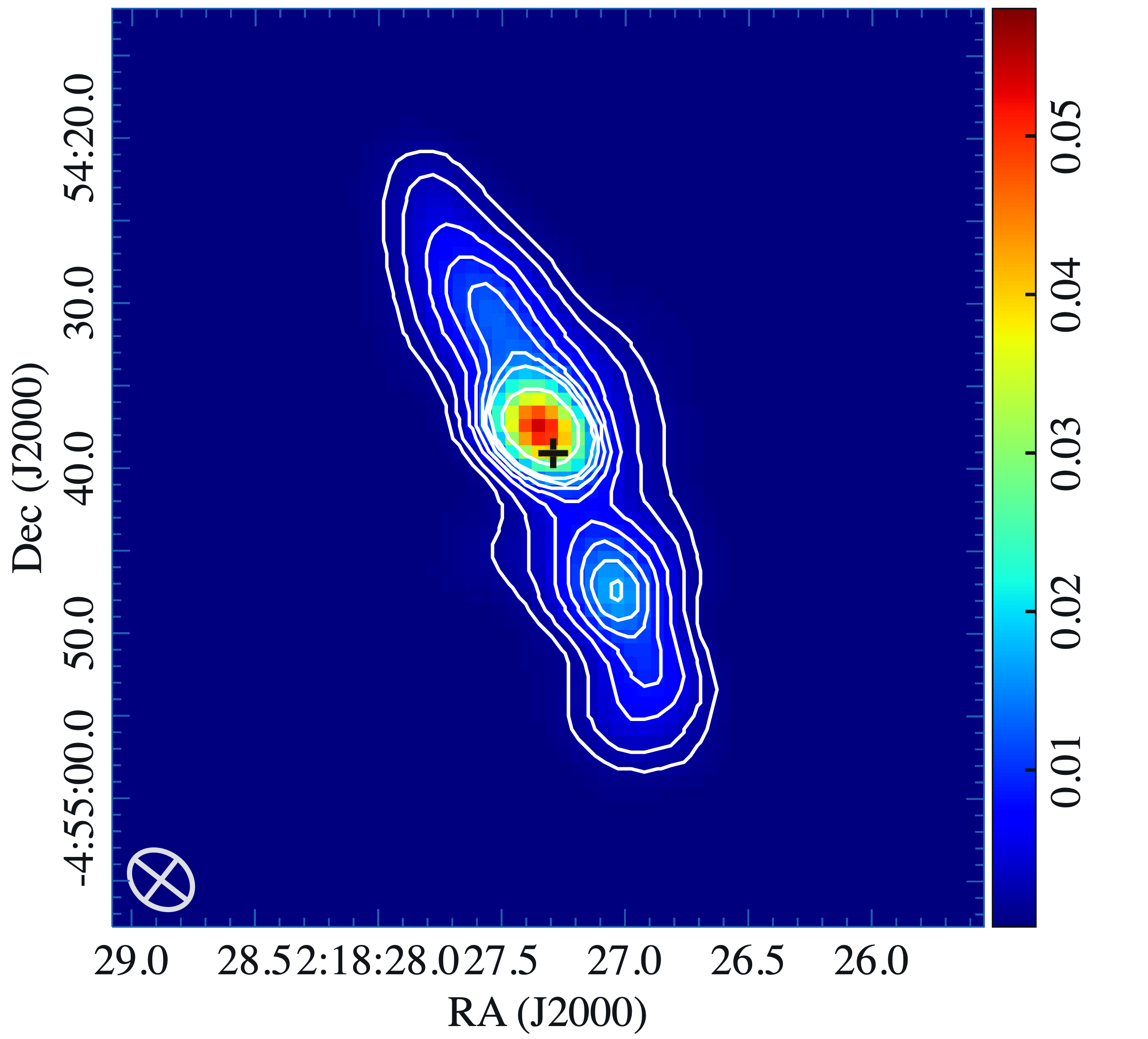}
\includegraphics[scale=0.16]{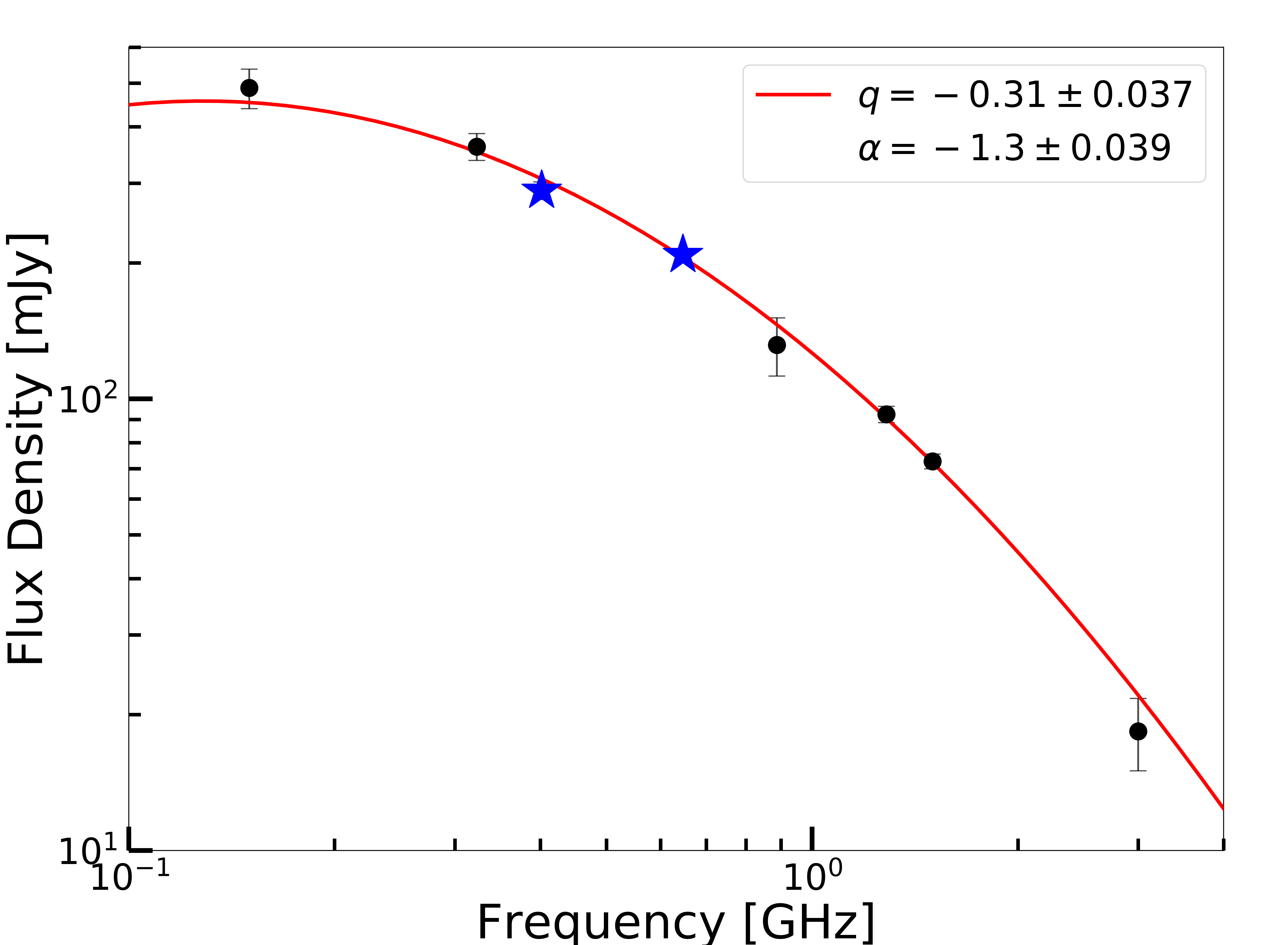}

\includegraphics[scale=0.05]{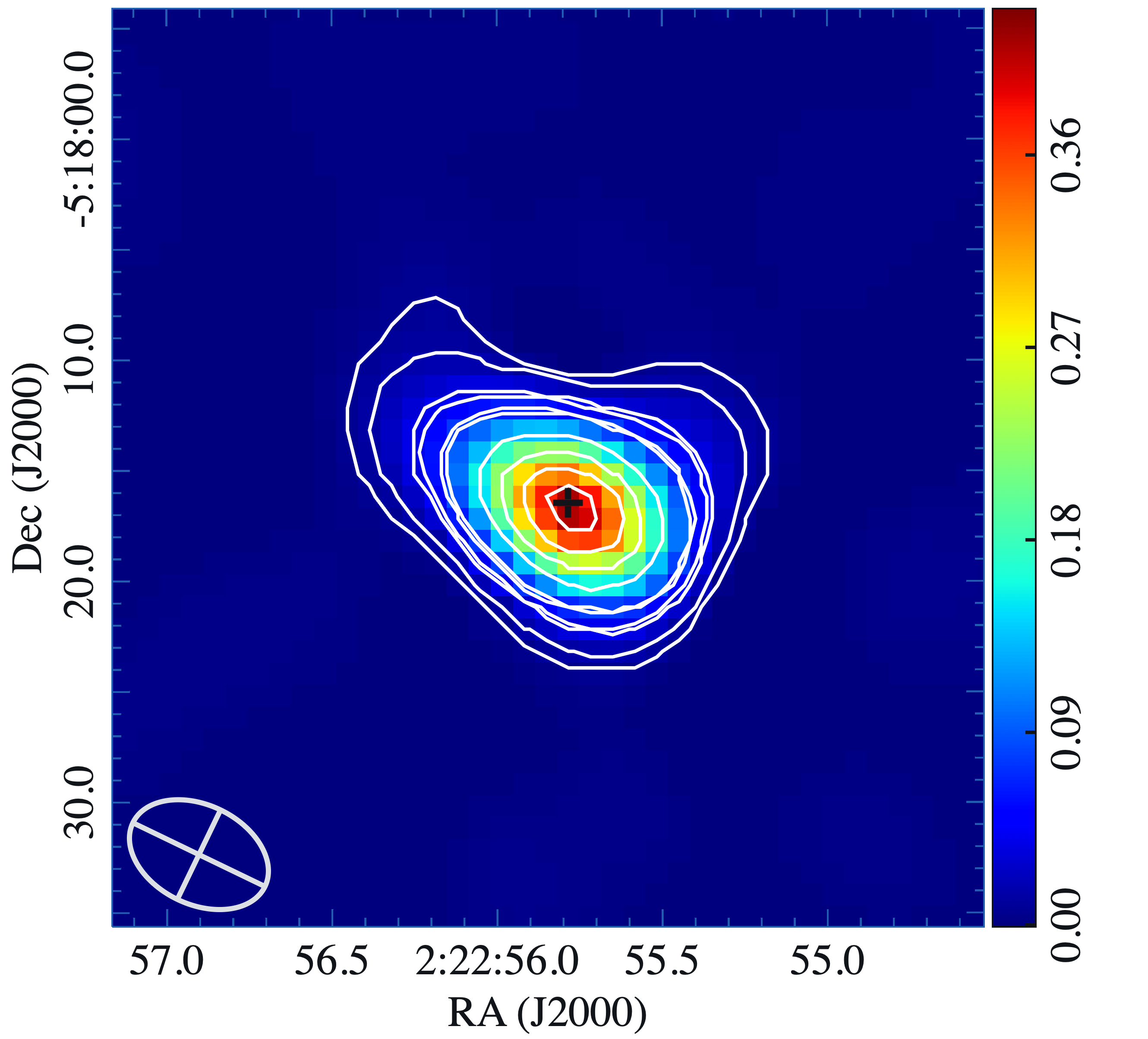}
\includegraphics[scale=0.05]{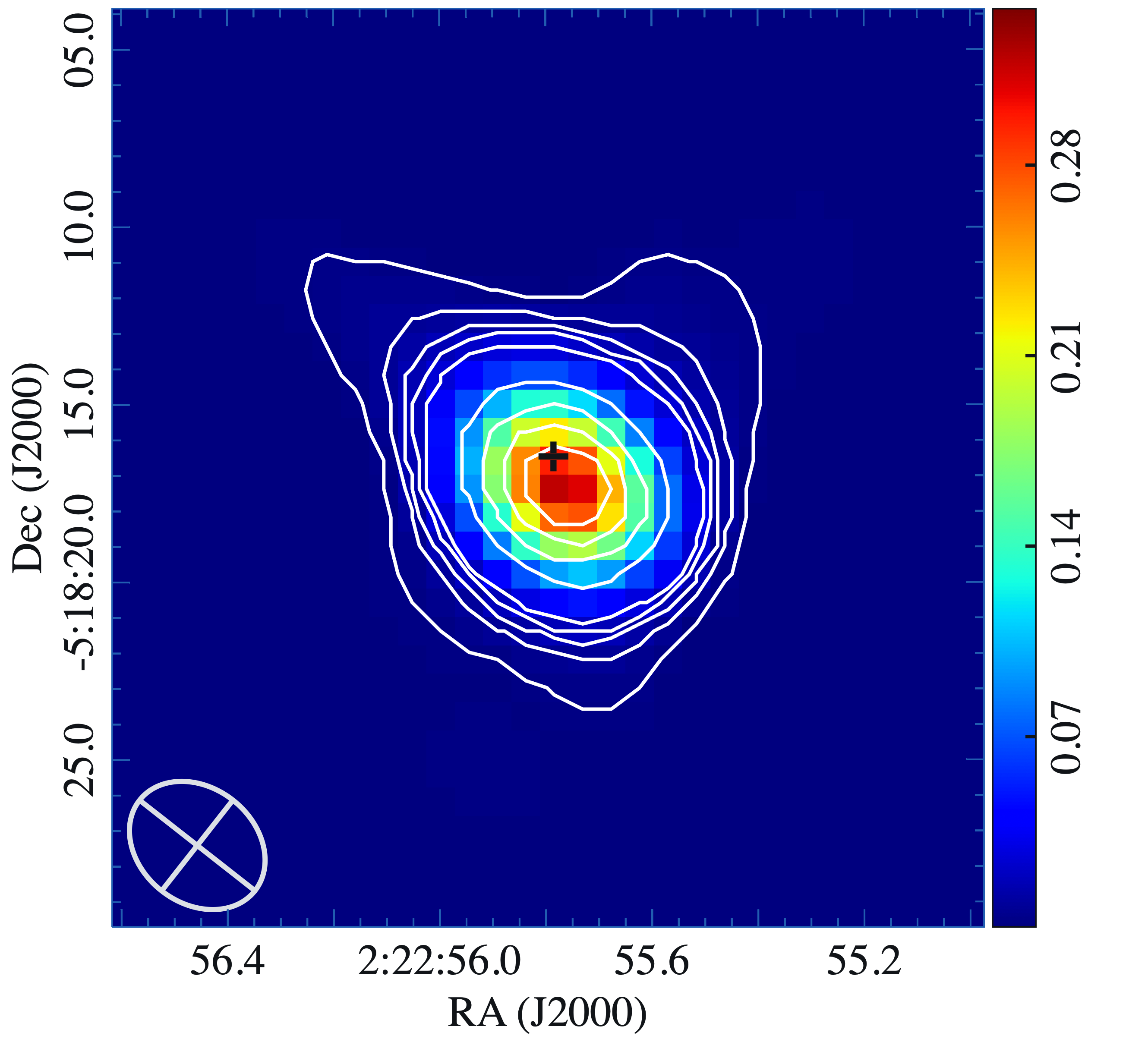}
\includegraphics[scale=0.16]{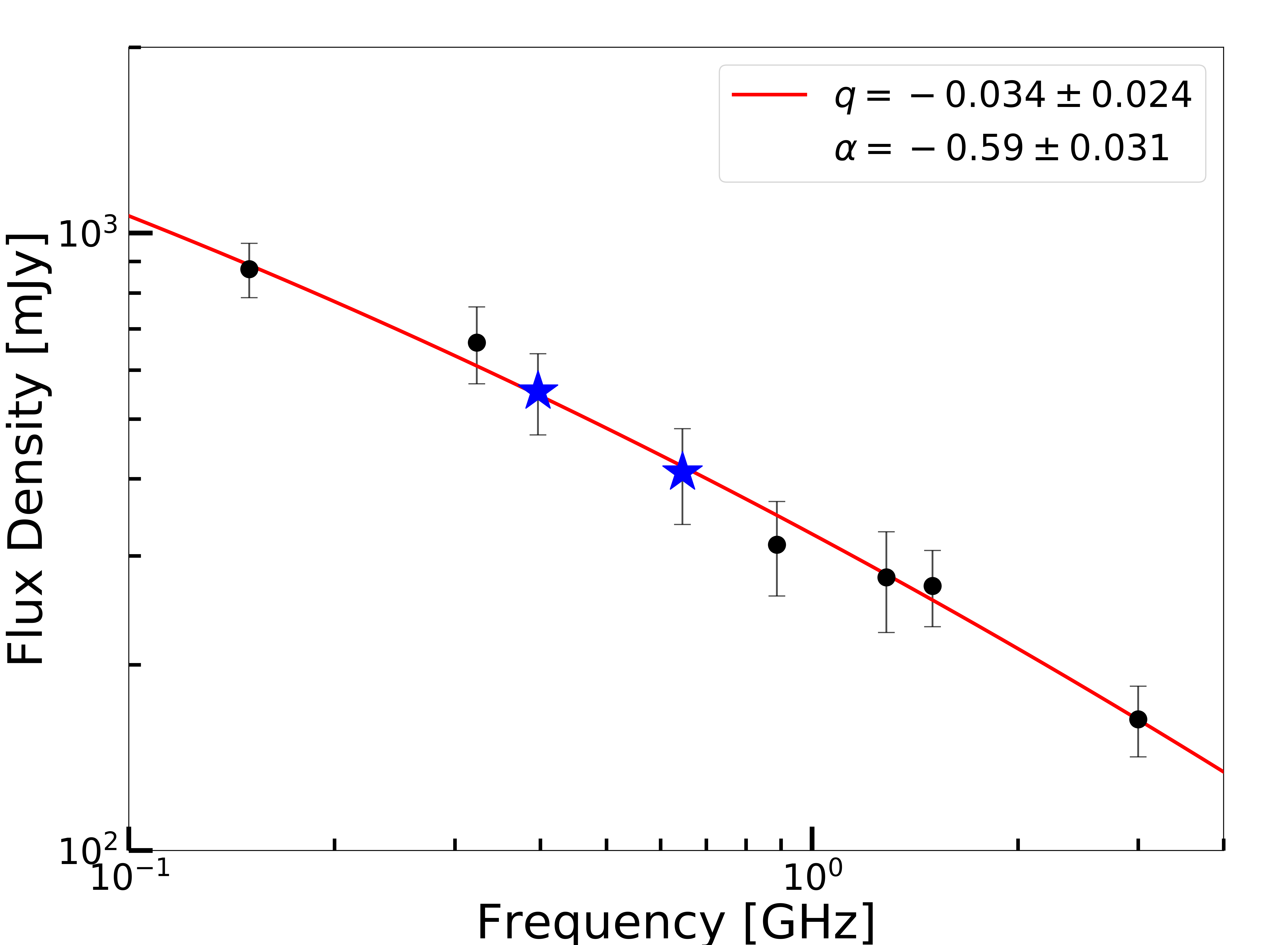}
\end{center}
\caption{Images of an extended source (upper panel), and a compact source (lower panel), at band~3 (left panel) and band~4 (middle panel) from the XMM-LSS robust = $-$0.4 images. The radio contour levels are at 3$\sigma$ $\times$ (1, 2, 4, 8 ...). The `+' mark in the images represent the optical host positions. The right panels show the radio SEDs with the best-fit model (red solid curve). The star symbols represent the band-3 and band-4 uGMRT data.}
\label{fig:SED1}
\end{figure*}

\begin{figure*}[ht]
\begin{center}
\includegraphics[width=17.5cm]{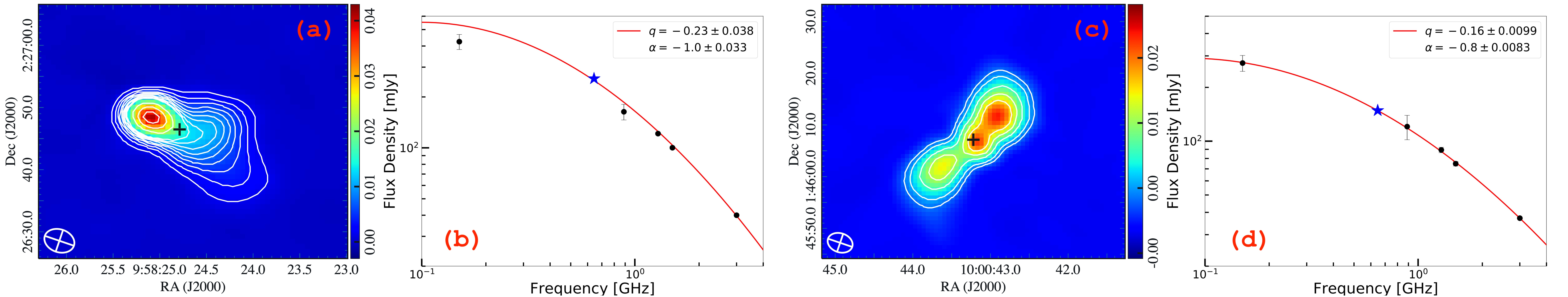} \\
\includegraphics[width=17.5cm]{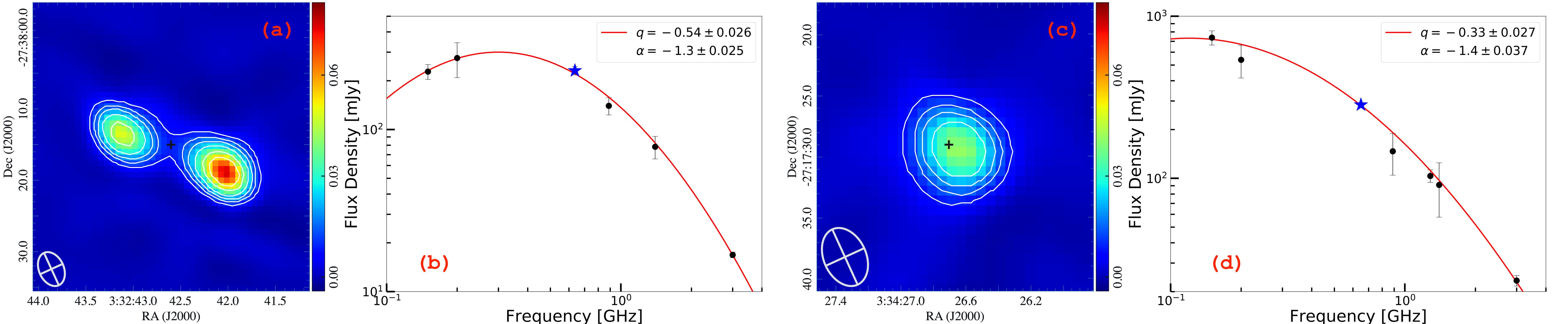}
\end{center}
\caption{Panels (a) and (c) show images of two bright sources at band~4 in the COSMOS (upper-panel images) and E-CDFS (lower-panel images) fields from the robust = $-$0.4 images. The radio contour levels and `+' symbol are as in Figure~\ref{fig:SED1}. Panels (b) and (d) show the corresponding radio SEDs, along with the best-fit models (red solid curve). The star-marked data point represents the superMIGHTEE band-4.}
\label{fig:SED-cosmos-ecdfs}
\end{figure*}

We used several radio sky surveys to construct the radio SEDs. 
The data points for these SEDs for the XMM-LSS field were obtained from the following surveys: 150~MHz TGSS ADR1 \citep[TIFR GMRT Sky Survey;][]{Intema17}, 323 MHz legacy GMRT \citep{Singh14}, band-3 and band-4 uGMRT (this work), 888 MHz RACS \citep[Rapid ASKAP Continuum Survey;][]{McConnell20}, 1284 MHz MeerKAT \citep[MIGHTEE DR1;][]{Hale25}, 1.5 GHz JVLA \citep{Heywood20}, and 3.0 GHz VLASS \citep[Very Large Array Sky Survey;][]{Lacy16}. 
We utilized the following datasets in the case of the COSMOS field: 150 MHz TGSS ADR1, band-4 uGMRT (this work), 888 MHz RACS, 1,284 MHz MeerKAT (MIGHTEE DR1), 1.4 GHz FIRST/NVSS \citep[FIRST: Faint Images of the Radio Sky at Twenty-cm; NVSS: NRAO VLA Sky Survey,][respectively]{Becker95,Condon98}, and 3.0 GHz VLASS.
Finally, we utilized seven datasets for the E-CDFS field, including the 150 MHz TGSS ADR1, the 200 MHz GLEAM \citep[GaLactic and Extragalactic All-sky MWA survey;][]{Wayth15}, the band-4 uGMRT dataset (this work), the 888 MHz RACS, the 1,284 MHz MIGHTEE DR1, the 1.4 GHz NVSS, and the 3.0 GHz VLASS. 

To illustrate, the images of two prominent (bright) radio sources in the XMM-LSS field$-$one extended and one pointlike$-$along with their corresponding SEDs are shown in Figure~\ref{fig:SED1} \citep[see also,][]{2023MNRAS.523..620P,Dutta23}.
Similarly, images and SEDs for sources in the COSMOS and E-CDFS fields are shown in Figure~\ref{fig:SED-cosmos-ecdfs}. The measurement of integrated flux densities for all images was conducted using the {\sc carta\footnote{\url{https://carta-beta.idia.ac.za}}} software by delineating a region that encompasses the sources of interest. This process also yields associated uncertainties (\textsc{rms}) in the flux densities of the sources. The data points utilized in the radio SEDs consist of these integrated flux density values obtained from {\sc carta}. The reduced $\chi^2$ values are close to 1 for all fitted radio SEDs, indicating a good fit. In all cases, the deviations of the superMIGHTEE flux density from the fitted curve are within the errors on the measurements. For example, in XMM-LSS, the deviations of the band~3 and band~4 uGMRT data points from the fitted curve are measured at 1.3\% and 2.3\%, respectively, for the J022255$-$051818 source, and 4.1\% and 1.3\%, respectively, for the J021827$-$045440 source (see Figure~\ref{fig:SED1}). Similarly, in the case of the COSMOS band-4 data, the deviation of the measurements from the fitted curve is 3.9\% for the J095822$+$022604 source and 1.2\% for the J100043$+$014609 source, and in the case of E-CDFS band-4 data, the deviations are 1.9\% for the J033427$-$271730 source and 1.7\% for the J033242$-$273818 source (see Figure~\ref{fig:SED-cosmos-ecdfs}). Thus, we believe that the systematic flux density errors for all the superMIGHTEE mosaic images are $<$5\%.

\section{Redshift Distributions}
\label{sec:redshift}

\begin{figure*}
   \begin{center}
\begin{tabular}{cccc}
    \includegraphics[width=0.495\columnwidth]{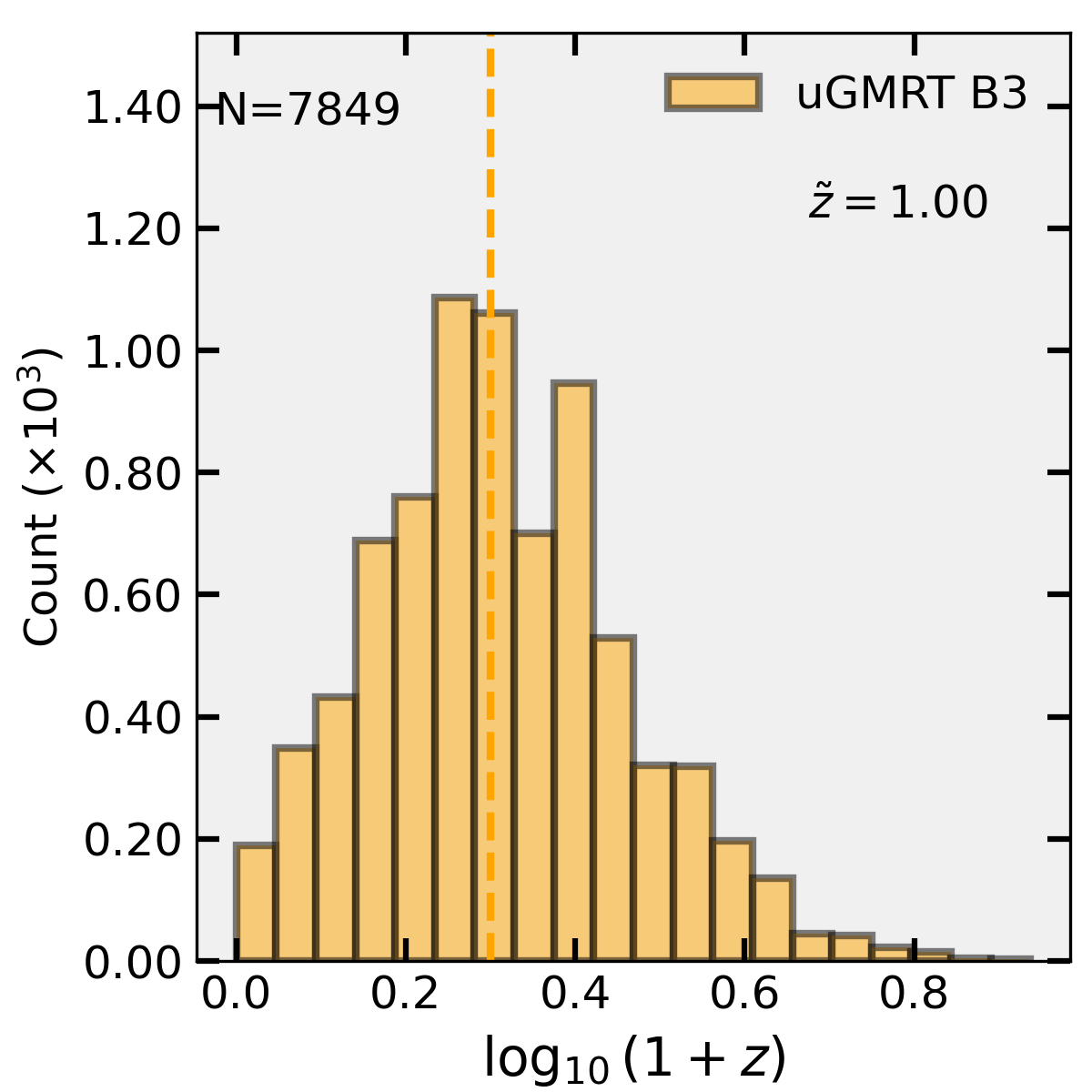} &
    \includegraphics[width=0.495\columnwidth]{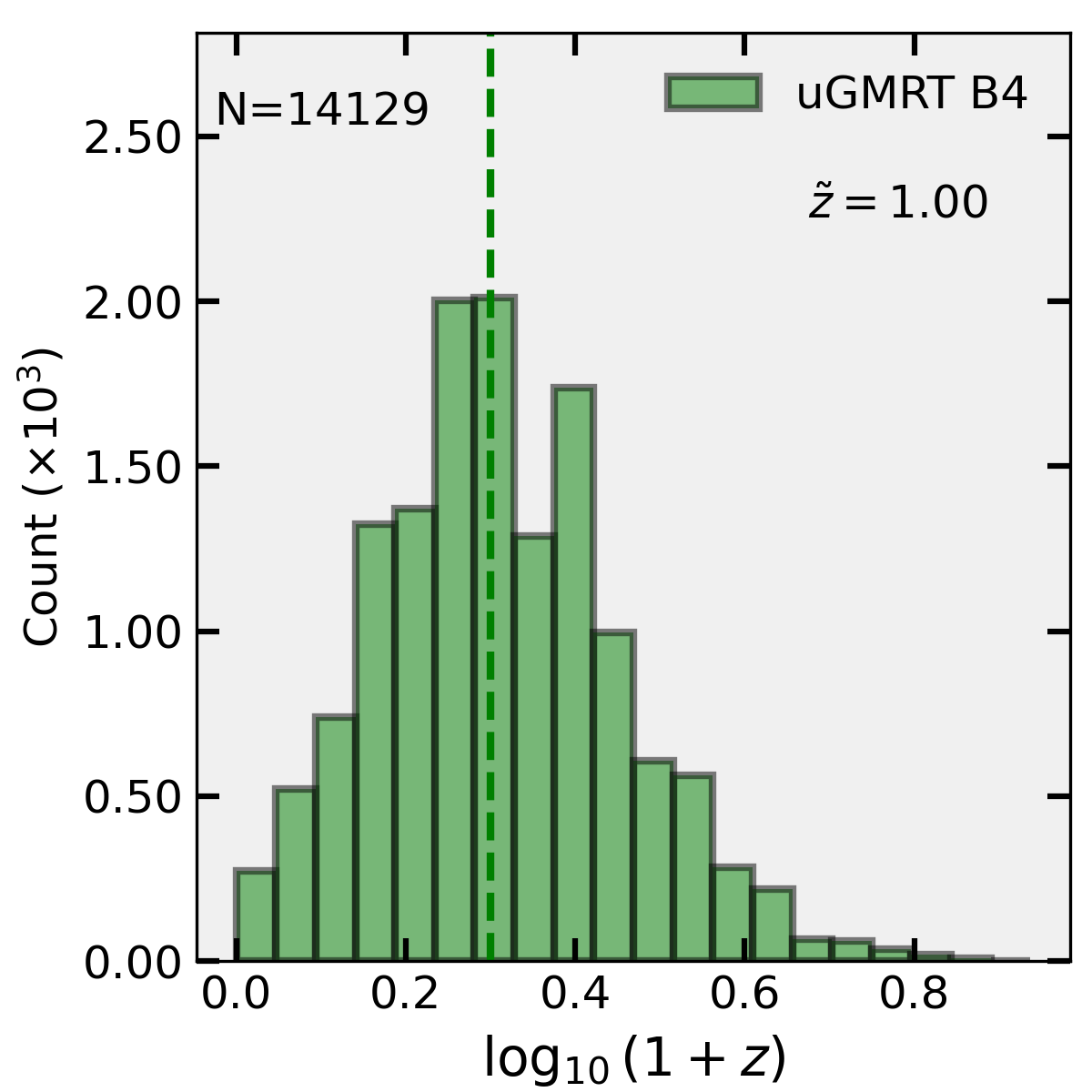} &
    \multicolumn{1}{c|}{\includegraphics[width=0.495\columnwidth]{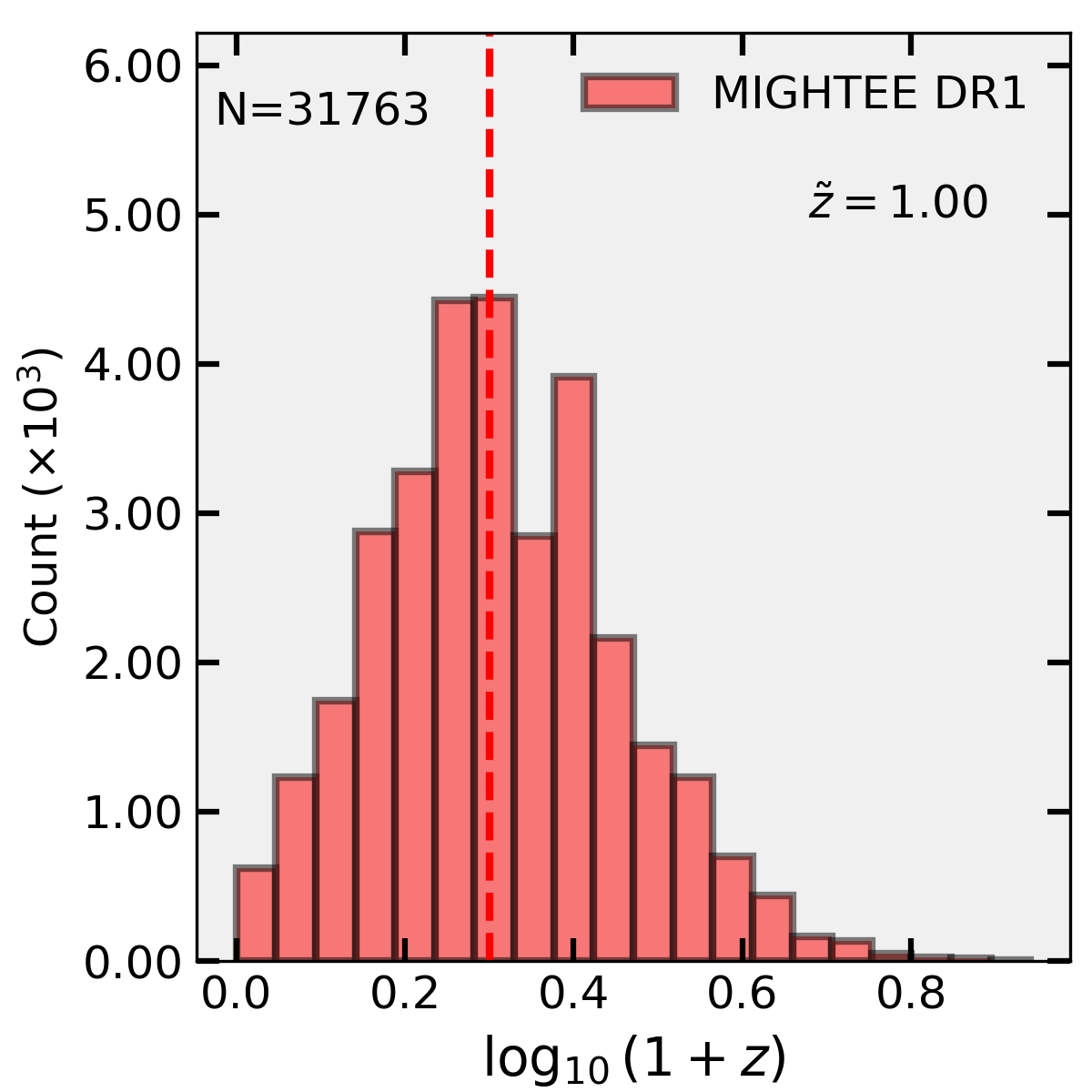}} & \\ \hline \\
    & \multicolumn{1}{c|}{} & & \\ [-0.9cm]
    \includegraphics[width=0.495\columnwidth]{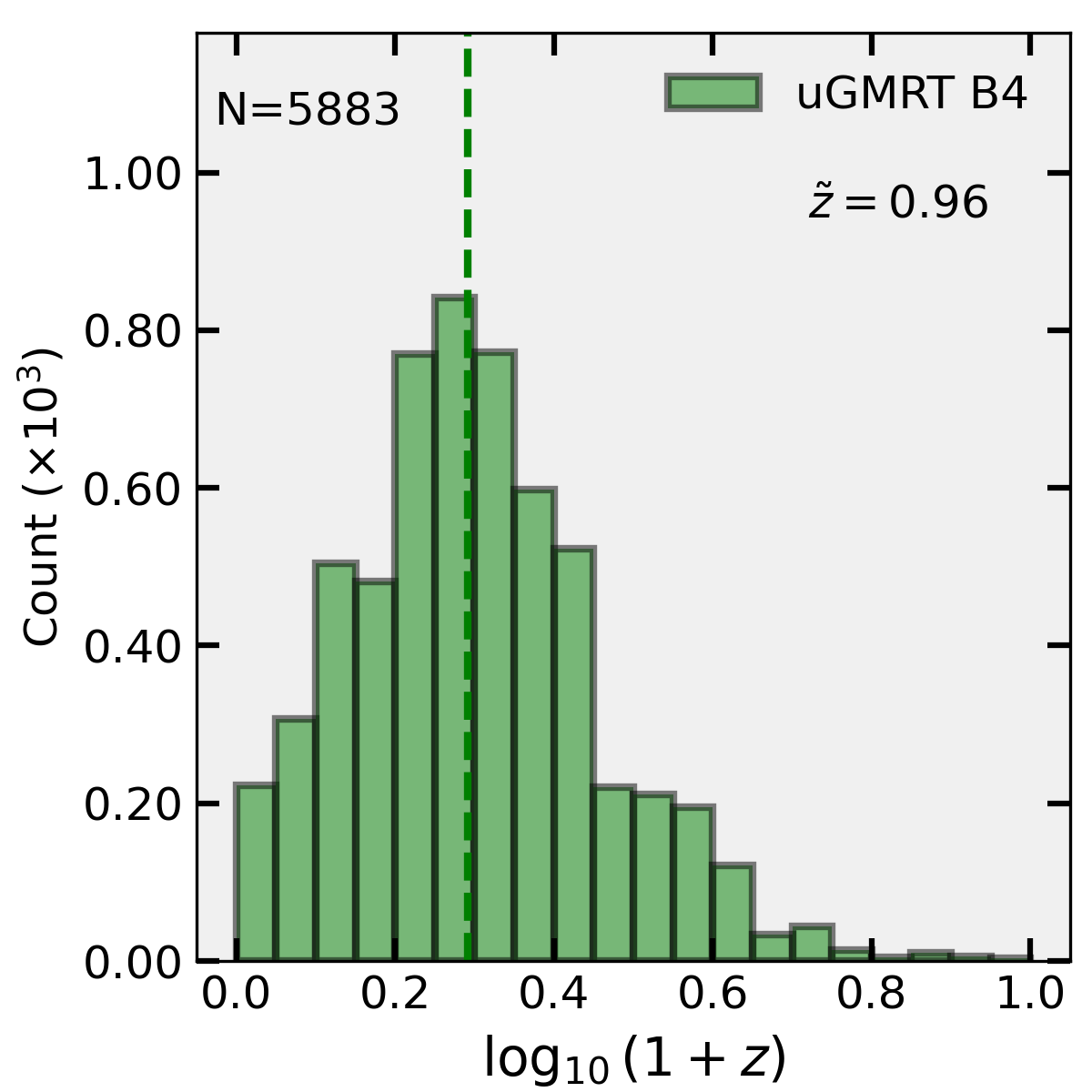} &
    \multicolumn{1}{c|}{\includegraphics[width=0.495\columnwidth]{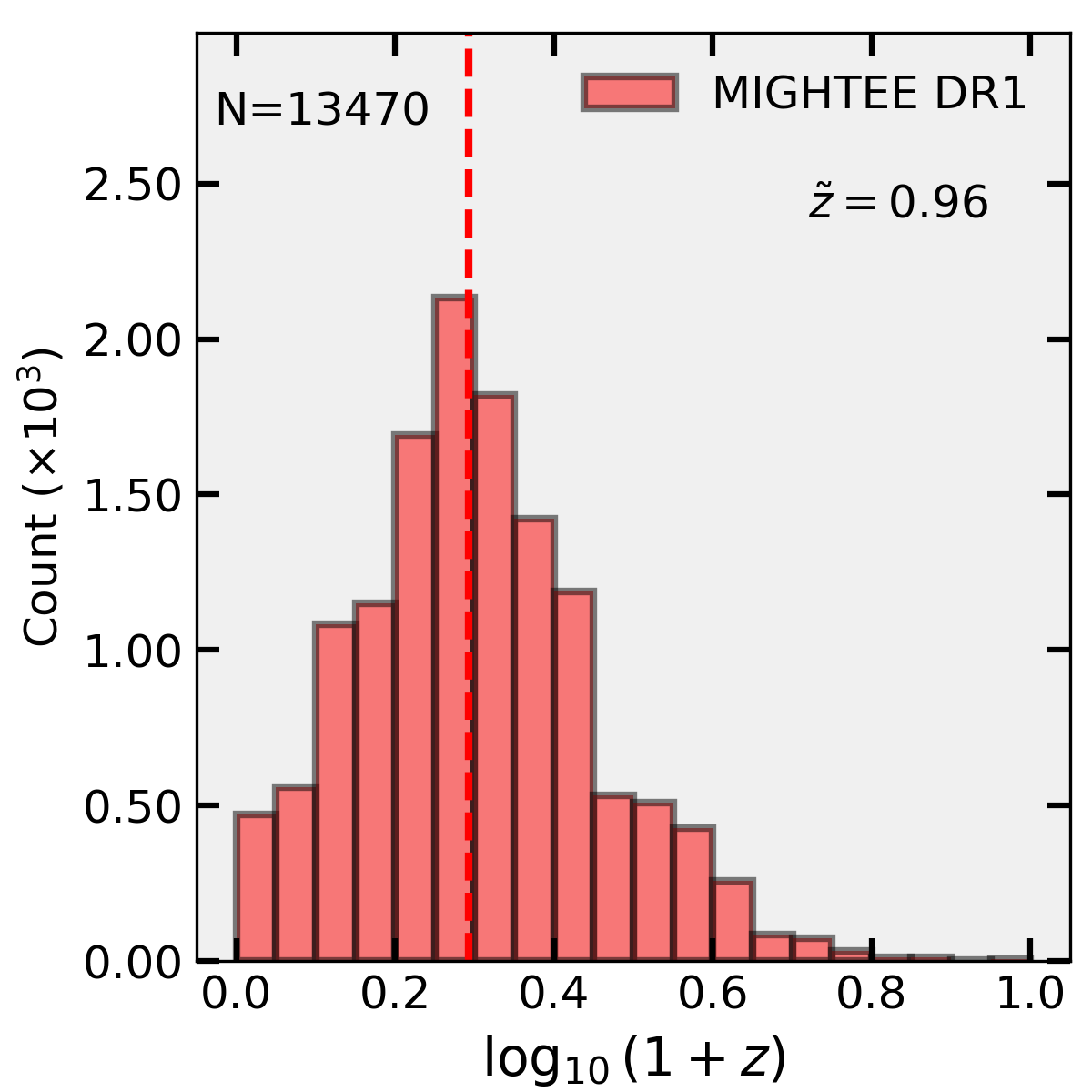}} &
    \includegraphics[width=0.495\columnwidth]{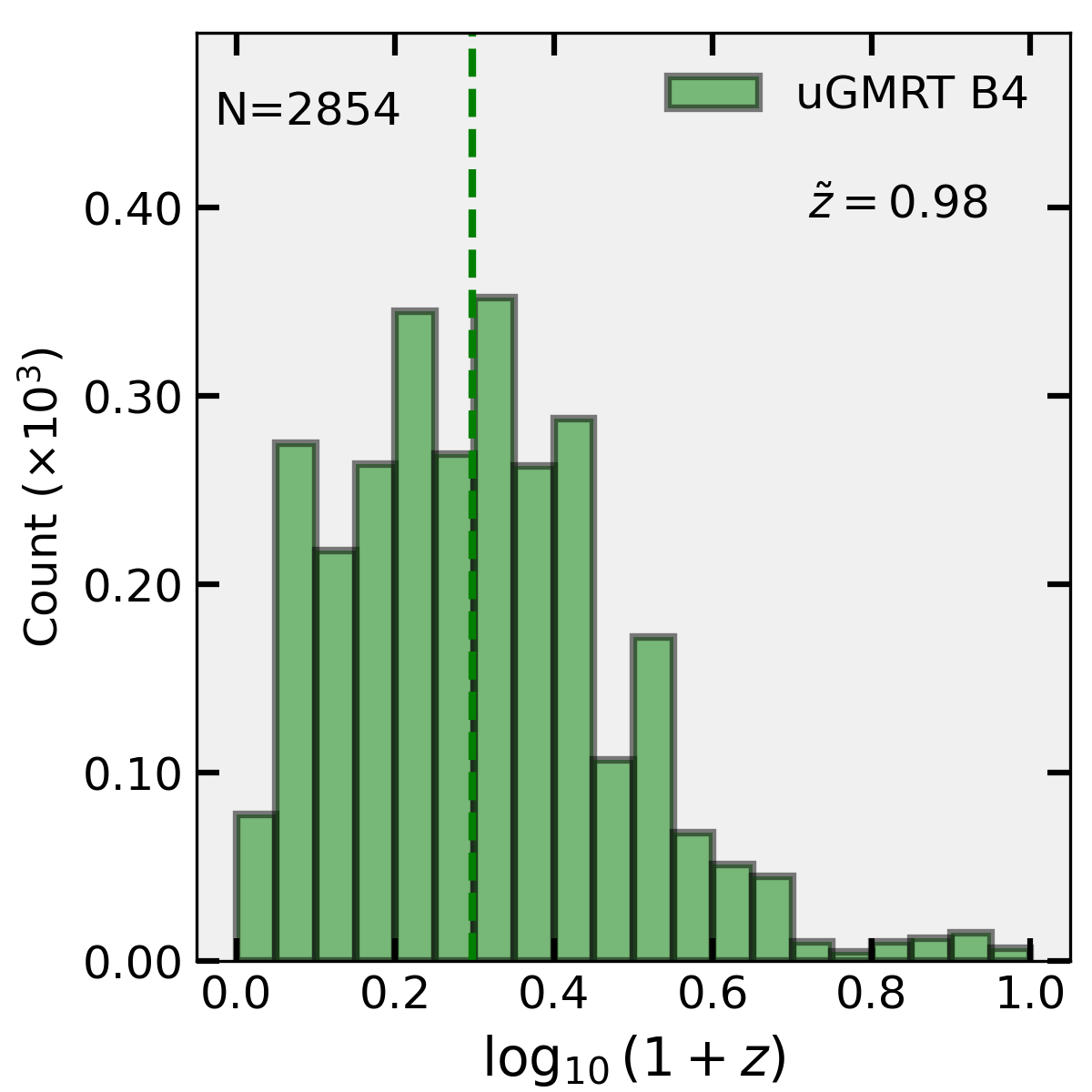} & 
    \includegraphics[width=0.495\columnwidth]{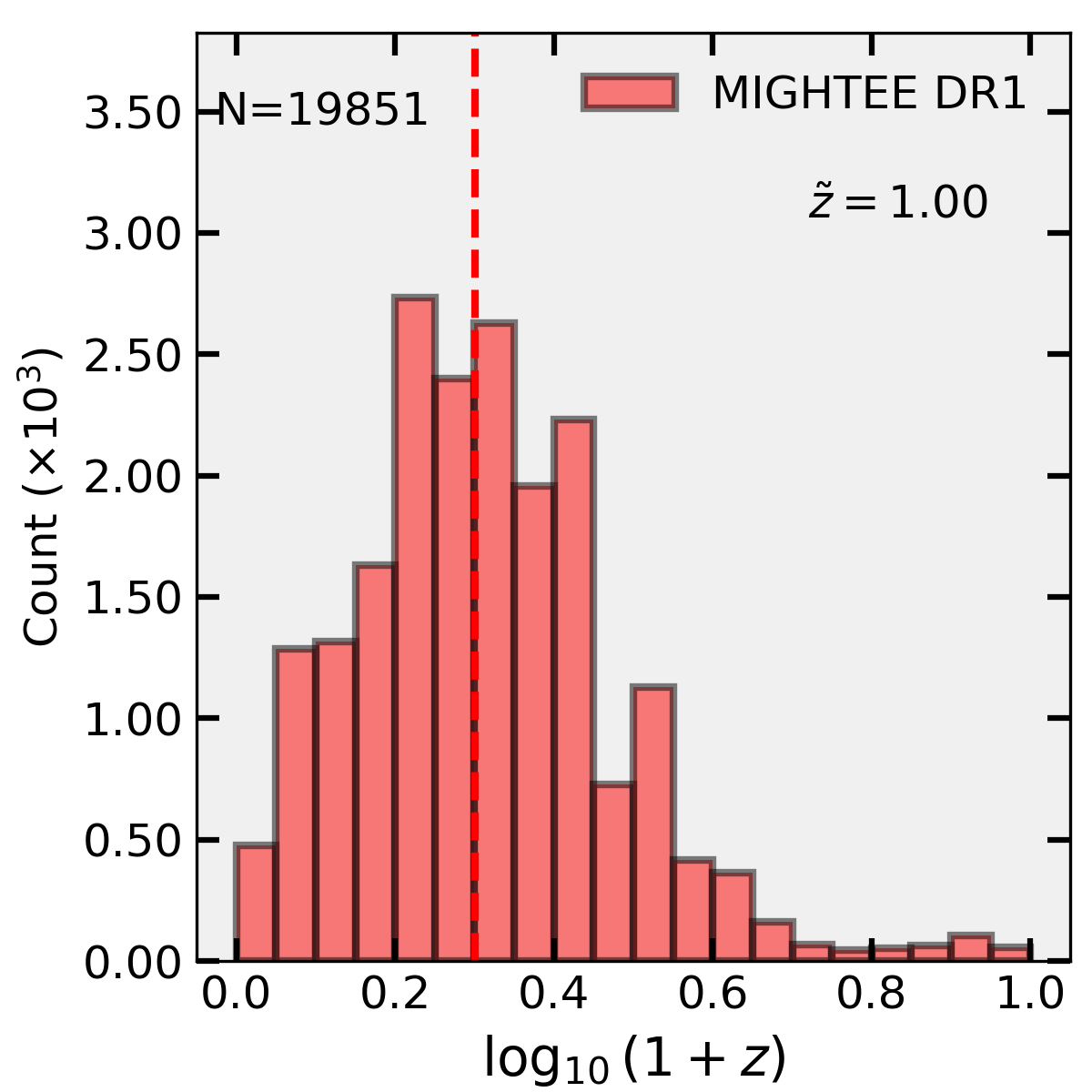}
\end{tabular}
   \end{center}
    \caption{Top row: the photometric redshift distributions for radio sources detected in uGMRT band~3 (left), band~4 (middle) and MIGHTEE DR1 (right) on the XMM-LSS field. Bottom row, left two panels: photometric redshift distributions for radio sources detected in uGMRT band~4 (bottom left) and MIGHTEE DR1 (bottom right) in the COSMOS field.  Bottom row, right two panels: photometric redshift distributions for radio sources detected in uGMRT band~4 (bottom left) and MIGHTEE DR1 (bottom right) in the E-CDFS field. In each distribution, the median redshift across the sample of radio sources is shown by the dashed line and listed below the legend along with the number of sources.}
    \label{fig:redshift-distributions}
\end{figure*}

The $YJHK_s$ photometric data are available for the XMM-LSS field from the VIDEO (VISTA Deep Extragalactic Observations) survey \citep{Jarvis13}, and COSMOS field from the {\it UltraVISTA} NIR imaging survey \citep{McCracken12}. The Hyper Suprime-Cam (HSC) Subaru Strategic Program (SSP) provides optical $GRIZy$ photometric detections \citep{Aihara18}. These optical/NIR photometry data are fed into a procedure that fits galaxy and AGN emission templates to the photometric data using {\sc LePhare} \citep{Arnouts11}. Additional photometric redshift estimations are derived using the GPz method machine learning procedure \citep{Hatfield22}. Based on the estimations from both these methods, a consensus photometric redshift is selected for each source using a hierarchical Bayesian procedure described in \citet{Duncan18}. In our work, we make use of the photometric redshifts derived by the MIGHTEE Collaboration using this procedure for sources in the COSMOS, XMM-LSS, and E-CDFS fields. The resulting optical/NIR photometric redshift catalogs for COSMOS and XMM-LSS were compiled and published by \citet{Bowler2020} while the E-CDFS catalog is available through private communication.

The superMIGHTEE radio source catalogs are crossmatched with these optical/NIR redshift catalogs using a simple pair-match method. The procedure finds a nearest optical/NIR counterpart to a radio source on the sky within a crossmatch radius of 1.12$^{\prime\prime}$, 1.13$^{\prime\prime}$, and 1.20$^{\prime\prime}$ for the XMM-LSS, COSMOS, and E-CDFS fields, respectively. The maximum crossmatch radius is determined by finding the intersection between two distributions. The first distribution represents all sky separations found between real radio and optical sources up to 4$^{\prime\prime}$. The second distribution represents all sky separations between mock radio and optical sources. The intersection of these two distributions indicates the approximate crossmatch radius beyond which true counterparts are unlikely to be selected.

It is true that some extended radio sources, e.g., jetted AGN, may be represented by multiple Gaussian components in the radio source catalog. This can result in a misidentification of optical/NIR counterparts to the radio sources. Only a robust crosscorrelation method that combines both the likelihood ratio and visual inspections (or citizen science efforts) would overcome this problem \citep[e.g.][]{2022MNRAS.516..245W}, and we therefore leave this task for future work. Since extended sources are generally rare, flagging and correcting for them in our radio catalogs will not significantly alter the redshift distributions shown in the Figure~\ref{fig:redshift-distributions}.

The optical/NIR catalogs also contain redshifts from the VIPERS \citep{Garilli2014} and UDSz \citep{Bradshaw2013} spectroscopic surveys. Spectroscopic redshifts are only available for 1\%, 3\%, and 2\% of all optical/NIR identified sources in XMM-LSS, COSMOS, and E-CDFS, respectively. In the absence of spectroscopic redshifts, we use the photometric redshift by default.  In Table \ref{table:supermightee-radio-optical-statistics}, statistics on the radio-optical crossmatching procedure are shown. The redshift distributions for uGMRT band~3, band~4, and MIGHTEE DR1 across all superMIGHTEE DR1 fields are illustrated in Figure~\ref{fig:redshift-distributions}. The distributions are similar across the superMIGHTEE radio bands and observed fields, with median values of $z\simeq1$ and a drop-off that extends to a maximum values of $z\simeq4$. We compared these redshift distributions to those detected in a similarly deep surveys, VLA-COSMOS 3\,GHz \citep{Smolcic2017} and LOFAR Two-metre Sky Survey Data Release 2 \citep[LoTSS-DR2:][]{Shimwell2022}; a common feature of both surveys, when photometric redshifts are included, is the characteristic peak at $z\sim1$ with a tail that diminishes to zero sources by $z\gtrsim5$.

\begin{table}[hb]
\centering
\caption{Radio and Optical/NIR Crossmatch Statistics for Each Field and Observing Band}
\label{table:supermightee-radio-optical-statistics}
\begin{tabular}{l l | c c}
\hline
Survey & Observing & Radio & Optical \\
       & Band & Sources & Crossmatches (\%) \\
\hline
XMM-LSS   &      &       &  \\
 & uGMRT B3 & 10,931 & 7,849 (71.8 \%) \\
 & uGMRT B4 & 16,284 & 14,129 (86.8 \%) \\
 & MIGHTEE & 72,187 & 31,763 (44.0 \%) \\
\hline
COSMOS   &      &       &  \\
 & uGMRT B4 & 7,442 & 5,883 (79.1 \%) \\
 & MIGHTEE & 20,886 & 13,470 (64.5 \%) \\
\hline
E-CDFS   &      &       &  \\
 & uGMRT B4 & 3,375 & 2,854 (84.6 \%) \\
 & MIGHTEE &  21,152 & 19,851 (93.8 \%) \\
\hline
\end{tabular}
\end{table}

\section{Spectral Properties of the $\mu$Jy Source Population}
\label{sec:spec-properties}

The primary goal of the superMIGHTEE project is to provide data for broadband spectral studies of low-frequency radio sources, reaching $\mu$Jy flux densities and arcsecond-scale resolution across frequencies ranging from a few hundred MHz to a few GHz.
Several detailed studies are underway and will be the subject of subsequent papers. As shown in the last column of Table~\ref{tab:mosaics}, we have cataloged approximately ten thousand compact source components with information at 400 and 650\,MHz, together with complementary data at similar resolution from the MIGHTEE project. Here, we provide a brief overview of the spectral properties.
For this analysis, we use the XMM-LSS source catalogs from the robust = $-$0.4 circular beam mosaics at band-4 (5.0$^{\prime\prime}$ resolution) and band-3 (6.9$^{\prime\prime}$ resolution). We crossmatched the band-4 catalog of 15,558 sources with the band-3 catalog and the MeerKAT MIGHTEE DR1 5.0$^{\prime\prime}$ resolution L-band catalog \citep{Hale_2024}.
There are 9,215 matches between band-4 and band-3, and 13,477 matches between band-4 and L-band. A total of 8,747 sources are matched across all three catalogs.

\begin{figure}[ht]
\begin{center}
\begin{tabular}{c}
\includegraphics[width=\columnwidth]{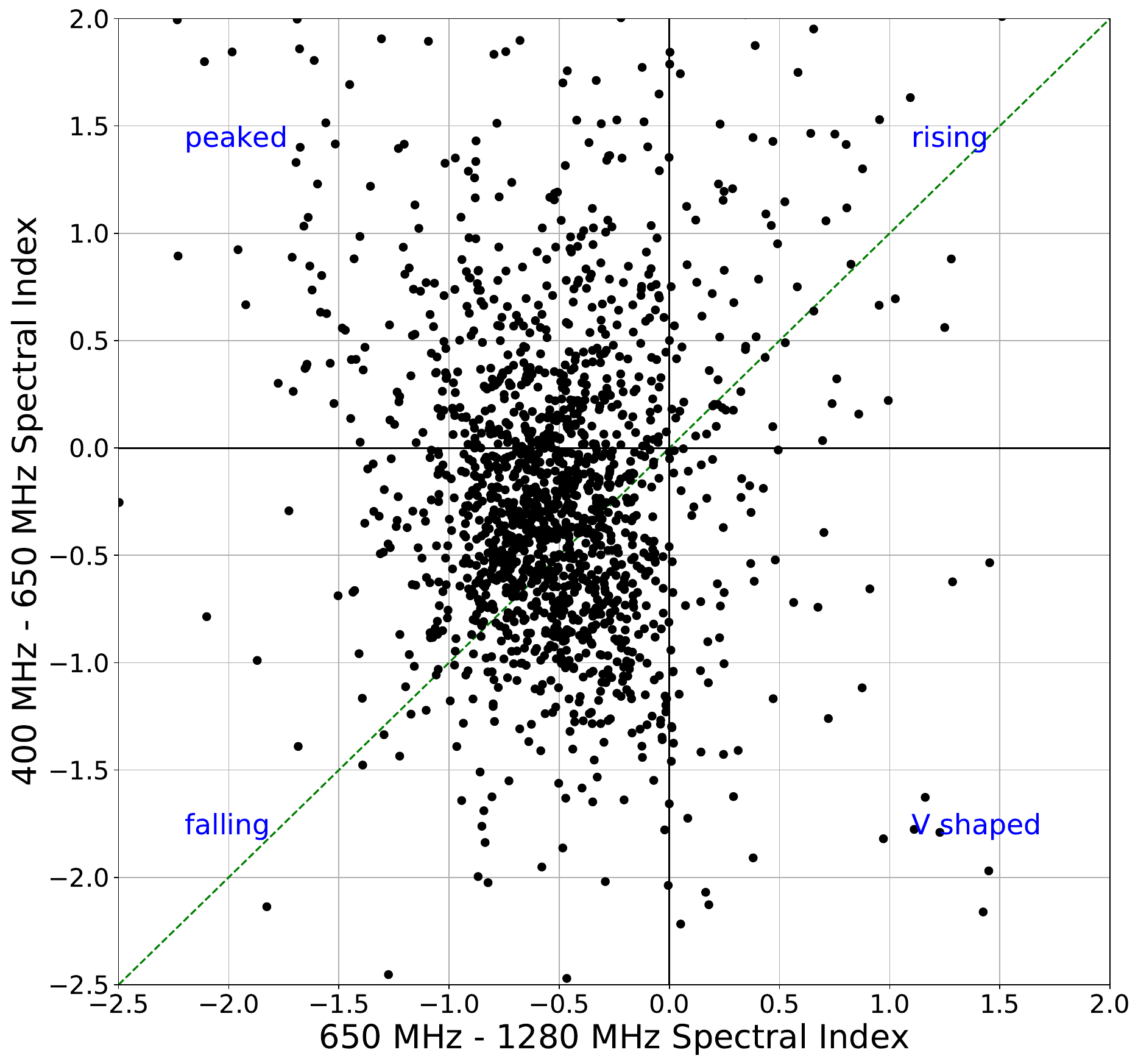} \\
\includegraphics[width=\columnwidth]{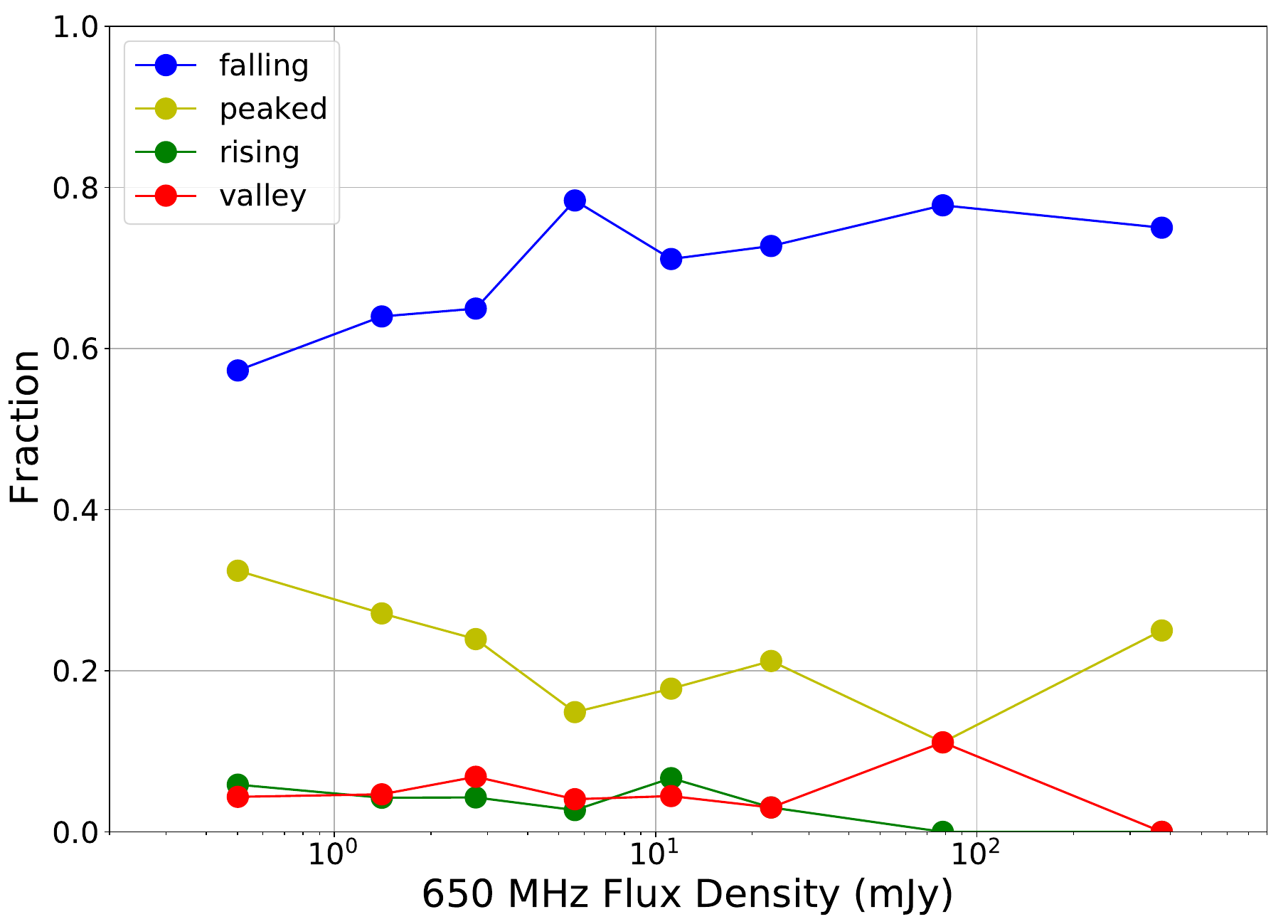} 
\end{tabular}
\end{center}
\caption{A color-color plot showing integrated flux density spectral index between 400 and 650 MHz versus the spectral index between 650 and 1280\,MHz for all detected sources from the XMM-LSS field with 650\,MHz integrated flux density above 400\,$\mu$Jy (upper panel).
The four quadrants of the plot are labeled for different spectral classes.
The fraction of the four types of color-color categories in Figure~\ref{fig:alpha-alpha} as a function of 650~MHz flux density (lower panel).}
\label{fig:alpha-alpha}
\end{figure}

Figure~\ref{fig:alpha-alpha} (upper panel) plots the spectral index of the cataloged integrated flux densities between 400 and 650\,MHz against the spectral index between 650 and 1280\,MHz from the XMM-LSS mosaics.
To avoid the effects of Eddington bias due to the higher noise at band~3, the plot is restricted to sources with a 650\,MHz total flux density greater than 300\,$\mu$Jy.
At this 650\,MHz flux density limit, 93.1\% of sources are also detected in the MIGHTEE catalog and 91.0\% are
detected in band 3.
A flat-spectrum object will have a signal-to-noise ratio greater than 8 at 400\,MHz.

The $\alpha$({400}--{650}) and $\alpha$({650}--{1280}) plot divides the sources into four quadrants:
(i) peaked sources, with the highest flux density at 650\,MHz;
(ii) falling sources, where the flux density monotonically decreases toward higher frequencies;
(iii) rising sources, where the flux density monotonically increases with frequency; and
(iv) V-shaped sources, where the lowest flux density occurs at 650\,MHz.

\begin{figure}[ht]
\begin{center}
\begin{tabular}{c}
\includegraphics[width=\columnwidth]{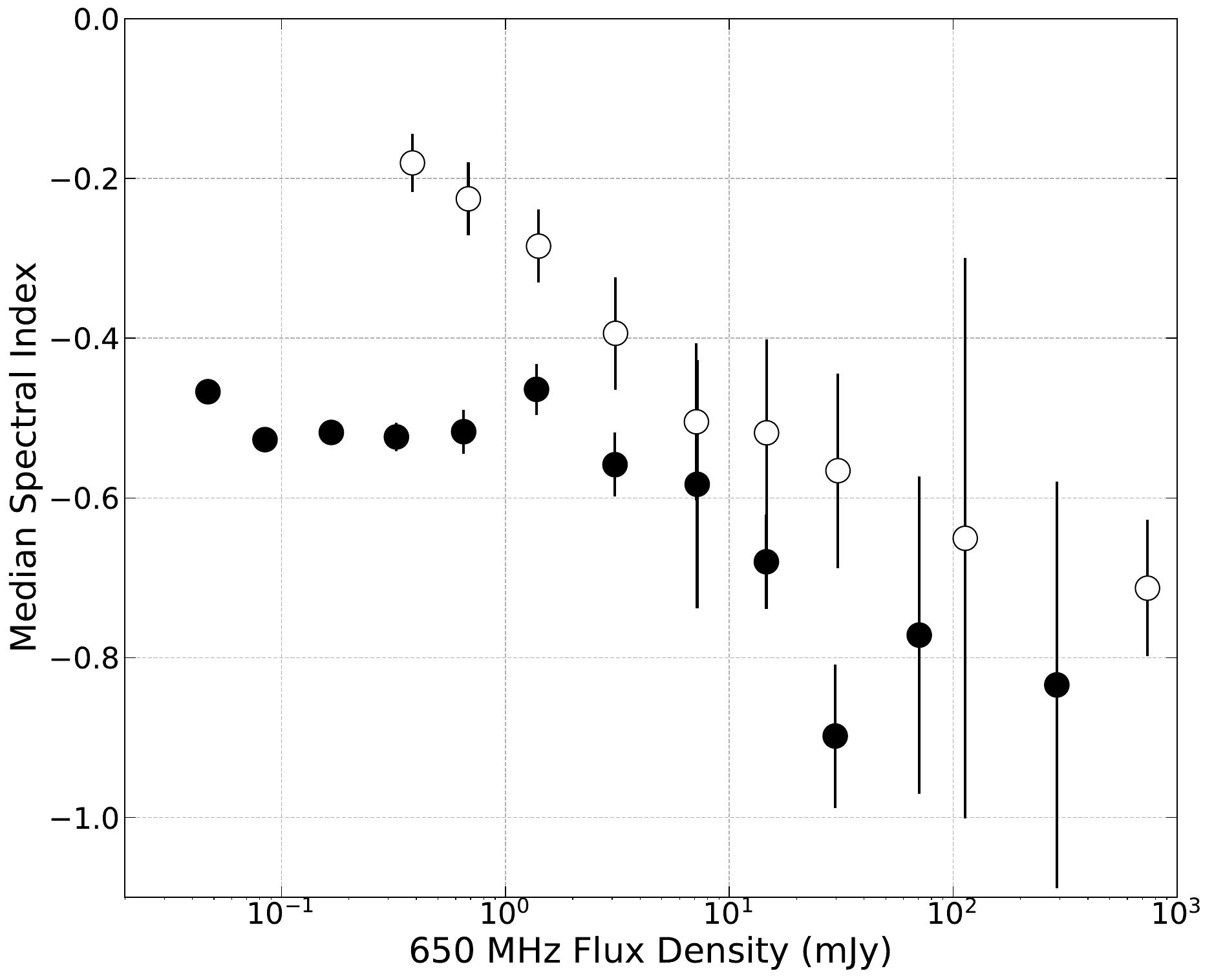} 
\end{tabular}
\end{center}
\caption{The plot shows the median 650--1280\,MHz spectral index (black circles) and the median 400--650\,MHz spectral index (white circles) as a function of 650\,MHz flux density.}
\label{fig:median-index}
\end{figure}

The vast majority of sources lie in the falling or peaked quadrants.
Of the 1,580 sources with integrated flux density greater than 400\,$\mu$Jy, 61\% have falling spectra, and 29\% have peaked spectra. Of the remainder, approximately equal fractions have rising and V-shaped spectra, 5.3\% and 4.6\% respectively.
This contrasts with a high flux density, high-frequency study by \cite{Murphy_2010}, based on the ATNF (Australia Telescope National Facility) 20\,GHz survey, which found 57\% falling spectra, 21\% peaked spectra, and 14\% rising spectra between 5, 8, and 20 GHz. In our low-frequency and low flux density data, there are more falling and peaked spectra and significantly fewer rising spectra.
As shown in Figure~\ref{fig:alpha-alpha} (lower panel), these fractions vary as a function of flux density.
As the 650~MHz flux density decreases, the fraction of sources with peaked spectra increases at the expense of the falling-spectrum sources. The fraction of falling spectra sources decreases from 75\% at high flux densities to $\sim$60\% at 500\,$\mu$Jy. Over the same flux density range, the peaked spectrum fraction increases from $\sim$20\% at high flux densities to $\sim$30\% at low flux densities. Both V-shaped and rising spectra remain approximately constant with the flux density at 4-5\%.

The median spectral indices $\alpha$({400}--{650}) and $\alpha$({650}--{1280}) as a function of 650\,MHz flux density are shown in Figure~\ref{fig:median-index}.
There are 9,215 sources with a 400--650\,MHz index and 13,477 sources with a 650--1280\,MHz index.
The figure shows a dramatic change in character at a transition point of $\sim$10\,mJy. Above this flux density, the median spectral index at both frequency ranges is approximately constant, with median values of
$\alpha$({400}--{650}) = $-0.57 \pm 0.09$ and
$\alpha$({650}--{1280}) = $-0.75 \pm 0.06$.
The data thus indicate that, on average, there is spectral flattening toward lower frequencies for sources above 10\,mJy, but no obvious trend with flux density between $\simeq$10\,mJy and a few hundred mJy.

Deep GMRT imaging at 610\,MHz \citep{2009MNRAS.397..281I, Ocran_2020} shows that the transition from a population dominated by AGN to one dominated by star-forming galaxies occurs at a 610\,MHz flux density of a few mJy \citep[see also,][]{2025MNRAS.537.3481P}.
The observed change in spectral properties can be attributed to this transition.
Below 10\,mJy, $\alpha$({650}--{1280}) increases, reaching a value slightly above $-0.6$.
It then remains approximately constant down to 100\,$\mu$Jy.
The median value below 10\,mJy is $-0.51 \pm 0.005$.
A study of the radio synchrotron spectra of 14 nearby galaxies observed with the Effelsberg Telescope by \cite{Klein_2018} found that the spectra were best represented by a low-frequency mean slope of $-0.59 \pm 0.2$, with a spectral break occurring between 1 and 12\,GHz.
This result is consistent with our median $\alpha$({650}--{1280}) index below 10\,mJy.
It is also in agreement with measurements of the integrated continuum spectra of a sample of 250 bright sources by \cite{Marvil_2015}, which showed a mean spectral index of $-0.55$ between 325\,MHz and 1.4\,GHz.

At lower frequencies, in contrast, Figure~\ref{fig:median-index} shows an increasing spectral flattening toward low flux densities.
There is a continued and pronounced increase in $\alpha$({650}--{1280}) below 10\,mJy, reaching a value of around $-0.2$ at a 650\,MHz flux density below 1\,mJy.
This trend toward an increasing low-frequency spectral index is consistent with the larger fraction of peaked sources at low flux densities, as seen in Figure~\ref{fig:alpha-alpha}.

\section{Discussion}
\label{sec:discussion}

The aim of the superMIGHTEE uGMRT observations, which complement the MeerKAT MIGHTEE survey, is to provide full-Stokes, high-sensitivity, high-resolution radio images across a frequency range from a few hundred MHz to a few GHz, at arcsecond resolution.
In this data release (DR1), an initial spectral analysis is carried out for several thousand sources with redshifts up to $z \sim 4$. The deep radio sample is dominated by star-forming galaxies, whose average spectral characteristics become evident at flux densities below a few mJy.

The $650$--$1280$ MHz spectral index for sources above 10\,mJy, measured as $-0.75 \pm 0.06$, is indicative of optically thin synchrotron emission from a power-law electron energy distribution. The steepening of the spectrum compared to the median $400$--$650$ MHz spectral index at these flux densities suggests that, on average, the spectral break from synchrotron losses occurs at $\sim$1~GHz.

The spectral index, $\alpha$({650}--{1280}) for sources below 10\,mJy remains optically thin, with a median value of $-0.51 \pm 0.005$. This is similar to the mean spectral index of $-0.50$ observed for Galactic supernova remnants \citep{Klein_2018}, suggesting that the electron energy injection spectrum for radio emission from the star-forming galaxy population is similar to that of supernovae.

The rapid spectral flattening observed toward lower flux densities at low frequencies (400--650 MHz) can also be attributed to the star-forming galaxy population. \cite{Ann_2024} studied the spectral indices of star-forming galaxies in the ELAIS N1 field between 150\,MHz and 5000\,MHz. The average spectral index for 94 objects detected with LOFAR at 150 MHz is $-0.46^{+0.03}_{-0.02}$. They found that the low-frequency spectrum flattens with increasing visual opacity $\tau_{\rm v}$, suggesting that free-free absorption becomes significant at lower frequencies.
A more detailed analysis of the astrophysical implications of these results, including the redshifts and classification of the radio sources, will be presented in subsequent papers.

\section{Summary}
\label{sec:summary}

This paper presents the initial superMIGHTEE DR1 of total intensity radio continuum images from observations conducted to date. Images and associated source catalogs are provided for a total of 9.9 deg$^2$ at 650 MHz and 6.9 deg$^2$ at 400 MHz in XMM-LSS, COSMOS, and E-CDFS. These images, along with the associated catalogs of 27,101 sources at 650 MHz and 10,946 sources at 400\,MHz, when combined with MeerKAT MIGHTEE data, provide unprecedented broadband radio photometry of the deep low-frequency sky down to $\mu$Jy flux density levels.
The project serves as a pathfinder for future combined SKA-mid and SKA-low investigations of the deep radio sky later this decade.

The population of radio sources includes objects with redshifts exceeding 4, but is dominated by sources in the range $z$ = 0.2--2. An overview of the spectral properties from 400\,MHz to 1280\,MHz reveals a distinct spectral transition that begins at $\sim$10\,mJy, marking the shift from AGN-dominated to star-forming galaxy-dominated radio sources. The $650-1280$ MHz spectra transition to a median spectral index of $-0.51$, consistent with optically thin emission from an electron energy distribution driven by supernovae. Fainter star-forming galaxies exhibit significant spectral flattening at low frequencies, which is indicative of increased radio opacity, likely due to free-free absorption by thermal electrons. More detailed investigations will be presented in subsequent papers.

The superMIGHTEE data products also include full-Stokes spectropolarimetric hypercubes, as well as high spectral resolution continuum-subtracted data for spectral line investigations. The radio polarimetry, in combination with MeerKAT MIGHTEE polarization data \citep{Taylor_2024}, will enable studies of cosmic magnetic fields through polarized emission from faint radio sources, potentially shedding light on the evolution of magnetic fields in early star-forming galaxies \cite[see, e.g.,][]{2020IAUGA..30..311S,2014ApJ...787...99S}.
The spectral line images will probe H\,I and OH emissions over redshift ranges of $0.50 \leq z_{\text{H\,I}} \leq 1.58$ and $0.75 \leq z_{\text{OH}} \leq 2.03$, respectively. While individual high-redshift H\,I detections will be rare, the extensive multiwavelength datasets enable stacking techniques for statistical searches of emission from distant galaxies \citep{2001A&A...372..768C}.
Finally, uGMRT band-2 observations of XMM-LSS are currently underway by the superMIGHTEE team to obtain data at 130--260 MHz. These datasets will be included in subsequent data releases from the superMIGHTEE team.

\section*{Acknowledgements}

We thank the anonymous referee for helpful comments that improved this paper.
D.V.L. and I.Ch. acknowledge the support of the Department of Atomic Energy, Government of India, under project No. 12-R\&D-TFR-5.02-0700.  S.S., S.D. and S.K. acknowledge financial support from the Inter-University Institute for Data Intensive Astronomy (IDIA). IDIA is a partnership of the University of Cape Town, the University of Pretoria, the University of the Western Cape.
We acknowledge the use of the \textsc{ilifu} cloud computing facility, a partnership between the University of Cape Town, the University of the Western Cape, the University of Stellenbosch, the Sol Plaatje University, the Cape Peninsula University of Technology, and the South African Radio Astronomy Observatory. The \textsc{ilifu} facility is supported by contributions from IDIA, the Computational Biology division at the University of Cape Town, and the Data Intensive Research Initiative of South Africa (DIRISA).
This work was carried out using the data processing pipelines developed at the Inter-University Institute for IDIA\footnote{\url{https://idia-pipelines.github.io}}. 
We thank the staff of the GMRT who made these observations possible.  The GMRT is run by the National Centre for Radio Astrophysics of the Tata Institute of Fundamental Research.
The MeerKAT telescope is operated by the South African Radio Astronomy Observatory, which is a facility of the National Research Foundation, an agency of the Department of Science and Innovation.
The superMIGHTEE project is part of the India-South Africa Flagship Program in Astronomy, funded by the Indian Department of Science and Technology and the South African National Research Foundation.

Facilities: {GMRT, MeerKAT}

Software: {CASA \citep{2022PASP..134k4501C}}, \textsc{carta}

\section*{Data Availability}

The GMRT data underlying this article are available via the GMRT online archive facility\footnote{\url{https://naps.ncra.tifr.res.in/goa/data/search}}. 
The MeerKAT data is publicly available via the SARAO archive\footnote{\url{https://archive.sarao.ac.za [archive.sarao.ac.za]}}.
The images and associated catalog data are available on the public data repository at the IDIA science gateway at \url{gateway.idia.ac.za} with DOI \url{http://dx.doi.org/10.71621/n9hd-sj09}.
All data analyses packages used in this work are publicly available, and their URLs have been noted in the main text.

\section*{ORCID iDs}

Dharam V. Lal \url{https://orcid.org/0000-0001-5470-305X}

Russ Taylor \url{https://orcid.org/0000-0001-9885-0676}

Srikrishna Sekhar \url{https://orcid.org/0000-0002-8418-9001}

Ch. Ishwara-Chandra \url{https://orcid.org/0000-0001-5356-1221}

Sushant Dutta  \url{https://orcid.org/0000-0002-6542-2939}

Sthabile Kolwa \url{https://orcid.org/0000-0001-9821-4987}

\bibliography{main-draft}{}
\bibliographystyle{aasjournalv7}

\end{document}